\documentclass[12pt]{article}
\usepackage{titlesec}
\usepackage{blindtext}
\usepackage{amsmath}
\usepackage{xcolor}
\usepackage{textgreek}
\usepackage{graphicx}
\usepackage{tikz}
\usepackage[square,numbers,sort&compress]{natbib}
\usepackage{url}
\usepackage{pgfgantt}
\usepackage[font=scriptsize]{caption} 
\usepackage{booktabs}
\usepackage{multirow}
\usepackage{subcaption}
\usepackage{overpic}
\usepackage{comment}
\usepackage{authblk}
\usepackage{float}

\title{The effects of fibre spatial distribution and relative orientation on the percolation and mechanics of stochastic fibre networks: A model of peptide hydrogels}
\date{}

\author[1]{Amir Hossein Namdar}
\author[2]{Nastaran Zoghi}
\author[3]{Aline Miller}
\author[4]{Alberto Saiani}
\author[1]{Tom Shearer}
\affil[1]{Department of Mathematics, School of Natural Sciences, Faculty of Science and Engineering, The University of Manchester, Oxford Road, M13 9PL, UK}
\affil[2]{Department of Materials and Manchester Institute of Biotechnology, School of Natural Sciences, Faculty of Science and Engineering, The University of Manchester, Oxford Road, M13 9PL, UK}
\affil[3]{Department of Chemical Engineering, School of Engineering, Faculty of Science and Engineering, The University of Manchester, Oxford Road, M13 9PL, UK}
\affil[4]{Division of Pharmacy and Optometry and Manchester Institute of Biotechnology, School of Health Sciences, Faculty of Biology, Medicine and Health, The University of Manchester, Oxford Road, M13 9PL, UK}

\begin{document}

\maketitle

\begin{abstract}
\noindent 
The structures of fibre networks can vary greatly due to fibre interactions during formation. We have modified the steps of generating Mikado networks to create two new model classes by altering the spatial distribution and relative orientation of their fibres to mimic the structures of self-assembling peptide hydrogels (SAPHs), whose physical properties depend strongly on their fibres' interactions. The results of our models and experiments on a set of beta-sheet forming SAPHs show that modifying a network's structure affects the percolation threshold and the mechanical behaviour of the material, both near percolation and at higher densities. 

\end{abstract}

\section{Introduction}

Many materials are constructed from fibre networks, from biological materials, such as collagen-based tissues and the cytoskeleton, to synthetic materials, such as gels and paper \cite{Picu_book}. These materials are often modelled as stochastic networks \cite{Head_2003,Lindstrom:2010ul,van_Oosten_2016,Casey_2021} to study their physical properties, such as percolation \cite{lij2009} and elastic modulus \cite{Head_2003,Wilhelm_2003}. It has been shown that, at densities near percolation, the network's elastic moduli vary proportionally to $(\rho-\rho_c)^f$, where $\rho$, $\rho_c$, and $f$ are the density ($\rho=N/L^2$, the number of fibres per unit square area of side length $L$), percolation threshold density, and elastic exponent of the network, respectively \cite{Stauffer_2018}. At higher densities ($\rho\gg\rho_c$), they vary proportionally to $\rho^{f_h}$, where $f_h$ is the high density elastic exponent \cite{Shahsavari_2013, gao2017}. 

Researchers have modified various aspects of fibre networks to investigate their material properties. \citet{Parzez2022} investigated network connectivity -- the mean number of fibre segments emerging from each node in the network (a segment is the portion of a fibre between two connected nodes). They showed that the elastic modulus in the bending dominated regime increases, but the exponent $f_h$ decreases, with increasing connectivity (average number of connections to other nodes that each network node has). The global orientation preference of fibres has also attracted attention \cite{Yuri2018,Ackermann2016}. \citet{Yuri2018}  modified fibre orientation with respect to a global coordinate system to create anisotropic networks and showed that they percolate at higher densities.

Network-based materials form in a wide variety of ways \cite{picu_2022-2}, which gives them dissimilar structures. For example, open cell foams differ from collagen gels. Foams are formed by introducing gas bubbles into monomer solution before solidification \cite{gibson_ashby_1997}, while collagen gels are formed by cells laying down fibrils which self-assemble \cite{Zhu-2018}. To model the former, it is more suitable to use Voronoi networks, and for the latter, Mikado networks \cite{Islam_2018}.

In this paper, we model self-assembling peptide hydrogels (SAPHs), a family of peptides extensively studied by our group for their self-assembly, gelation, and potential biomedical uses such as tissue engineering and drug delivery \cite{bolan-2024,elsawy-2022,lingard-2024}. In SAPHs, fibres formed of short peptide sequences create a network through self-association \cite{elsawy2016}. The fibres are sufficiently thick and rigid to ignore thermal fluctuations. Variations in amino acid sequences or environmental conditions can significantly alter fibre interactions, affecting the network's structure and mechanical properties, with elastic exponents ranging from 2 to 4.6 \cite{gao2017, Wychowaniec-2020}. Despite their importance, however, the effects of local fibre interactions on the structure and mechanical behaviour of the network are not fully understood. To understand how the structural changes caused by fibre interactions lead to mechanical changes, we modified the Mikado network generation process (we did not \textit{directly} model fibre formation, branching, interaction, filament entanglements, and thermal fluctuations) to vary the fibres' spatial distribution and local orientations relative to each other. Our results can be applied to any materials that are discovered, or developed, to have similar features.

\section{Model}

Two dimensional Mikado networks are created by randomly and independently placing a given number of fibres of length $l$ (we used $l=1$) with random orientation within a square domain of side length $L$.
We modified this process by changing two fibre features, their distribution in space and their relative orientation, as follows:
\begin{itemize}
\item To modify spatial distribution, after randomly determining a position for a fibre's centre, a probability, $P_e$, was assigned based on the number of existing fibre centres within a radius, $r_e$. A random number between 0 and 1 was then generated, and if it exceeded $P_e$, the fibre was deleted.

\item To modify orientation, the orientations of fibres within a radius, $r_\theta$, were used to define the distribution from which the candidate fibre's orientation was drawn, increasing the likelihood of a specified alignment relative to existing fibres.

\end{itemize}

The defined probabilities and the critical radii can be specified in various forms. The forms we chose are given below. For the existence probability, we specified 
\begin{equation} \label{eq1}
P_e = (1-P_r)^{n_{e}},
\end{equation}
\begin{equation} \label{eq2}
r_e=\alpha\sqrt{\frac{1}{\rho}},
\end{equation}
where $P_r$, $n_e$, and $\alpha$ are, respectively: the reduction in the probability of the new fibre existing due to the presence of a single fibre centre previously placed within the circle, the number of fibre centres that already exist within that circle, and a constant which scales $r_e$ relative to $\sqrt{1/\rho}$, the distance between nodes in perfect square lattice. In our simulations we chose $\alpha=4/3$, which maximises the homogeneity of the fibres' spatial distibution (for smaller and larger values, the fibre density is less homogeneous). The orientation is chosen, based on the orientations of the existing fibres within a circle of radius $r_\theta$ (we chose $r_\theta=l/2$, since this is the maximal distance between two fibre centres such that they can theoretically be connected by a fibre), from a Gaussian mixture distribution \cite{Deisenroth_Faisal_Ong_2020}:
\begin{equation} \label{eq5}
g(\theta)=\sum_{i=1}^{n_{\theta}} \frac{1}{n_\theta} \frac{1}{\sigma\sqrt{2\pi}}\exp{\left(-\frac{(\theta-(\theta_{fi} + \Theta))^2}{2\sigma^2}\right)},
\end{equation}
where $\Theta$, $n_{\theta}$, $\sigma$, and $\theta_{fi}$, are the preferred orientation relative to the existing fibres, the number of fibres within the circle, the standard deviation of the distribution (we used $\sigma=1/\sqrt{12}$, a value small enough to ensure the angle of the new fibre is sufficiently affected by existing fibres), and the angle the $i^\text{th}$ fibre makes with the $x$-axis, respectively. 

\begin{figure}
   \centering
     \begin{subfigure}[b]{0.32\textwidth}
         \centering
         \begin{overpic}[width=1\textwidth]{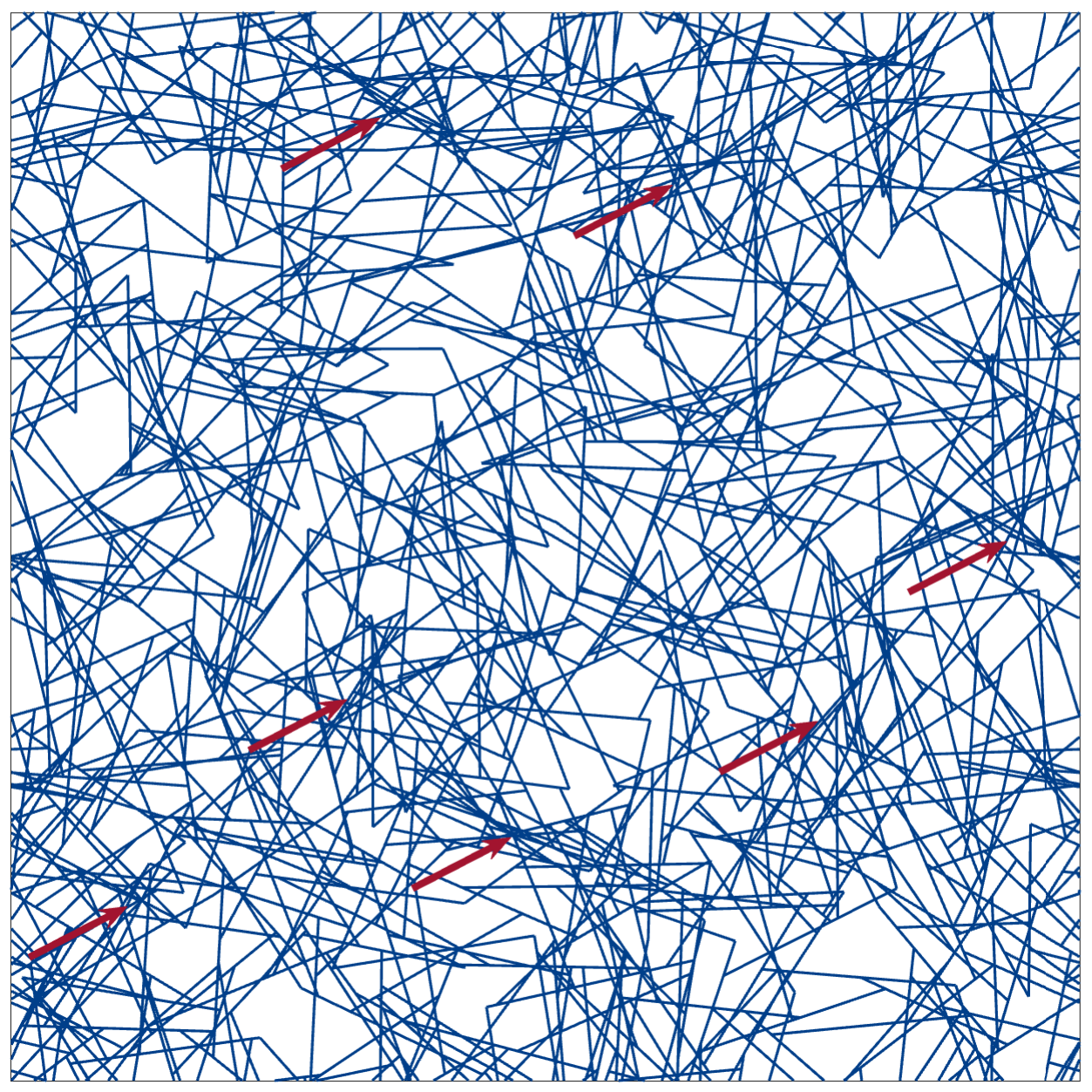}
         \put(5, 5){\colorbox{white}{\textbf{Mikado}}}
         \end{overpic}
     \end{subfigure}
     \hfill
     \begin{subfigure}[b]{0.32\textwidth}
         \begin{overpic}[width=1\textwidth]{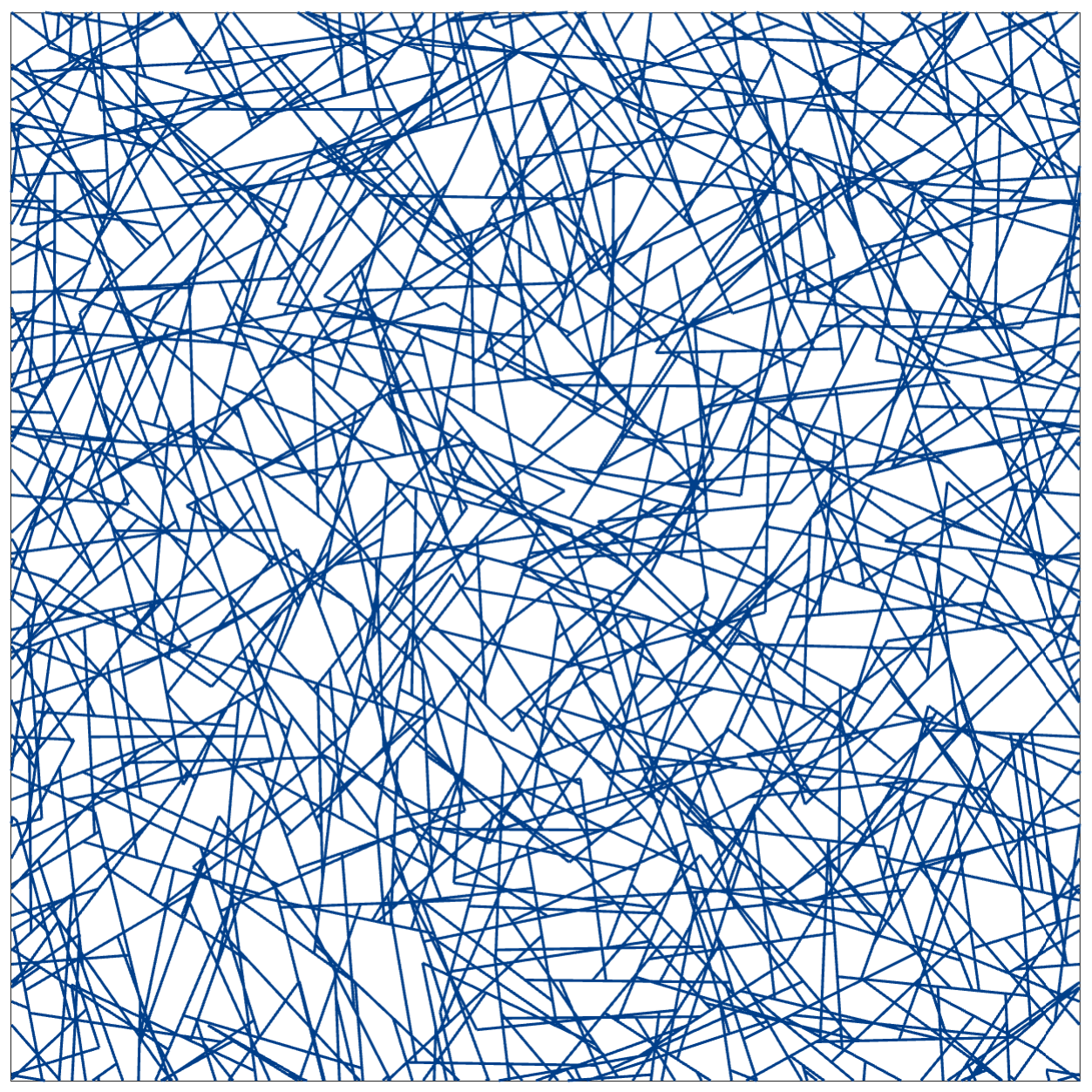}
         \put(5, 5){\colorbox{white}{\textbf{DM}}}
         \end{overpic}
     \end{subfigure}
     \hfill 
     \begin{subfigure}[b]{0.32\textwidth}
         \centering
         \begin{overpic}[width=1\textwidth]{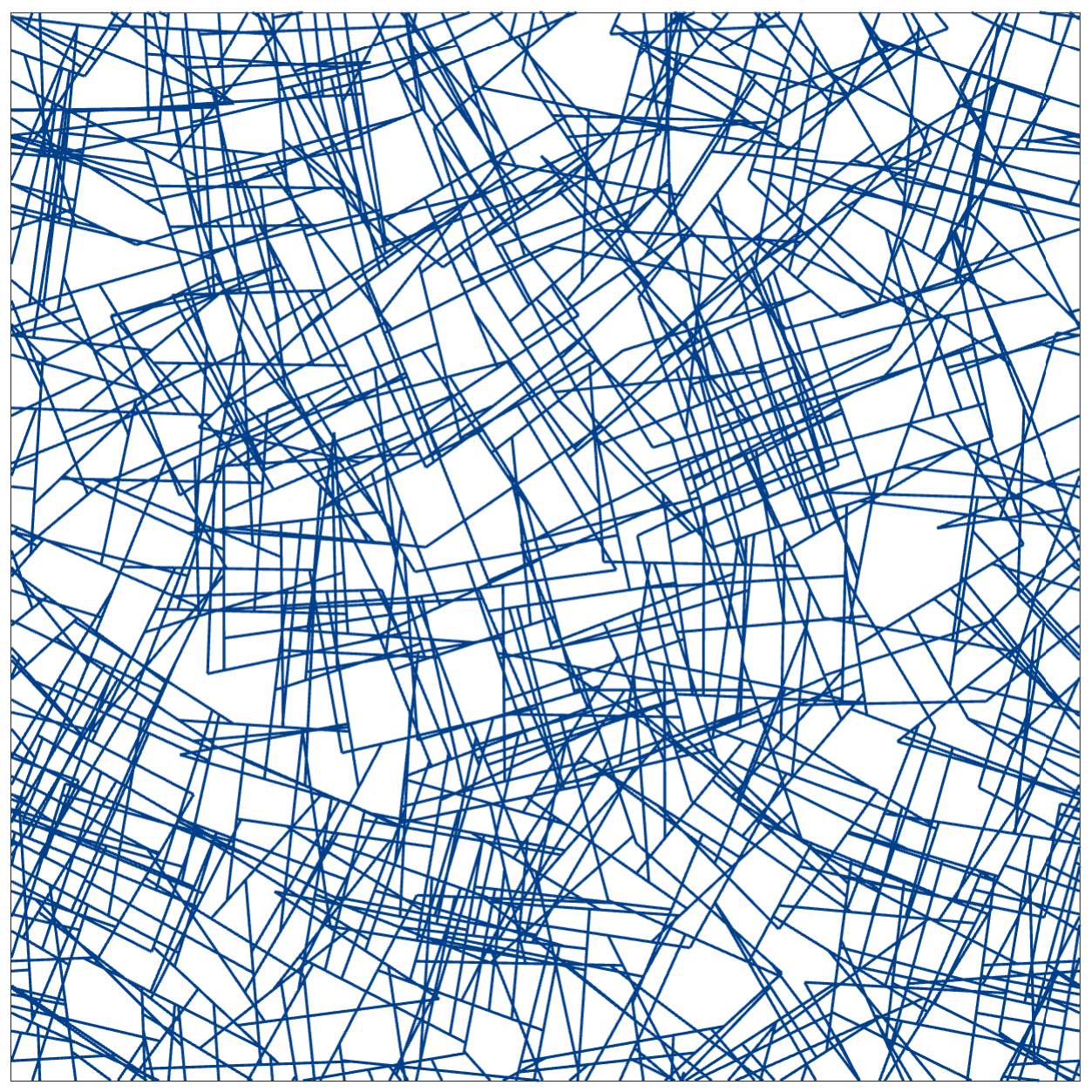}
         \put(5, 5){\colorbox{white}{\textbf{DOM type1}}}
         \end{overpic}
     \end{subfigure}
     \hfill \\
      \begin{subfigure}[b]{0.32\textwidth}
         \centering
         \begin{overpic}[width=1\textwidth]{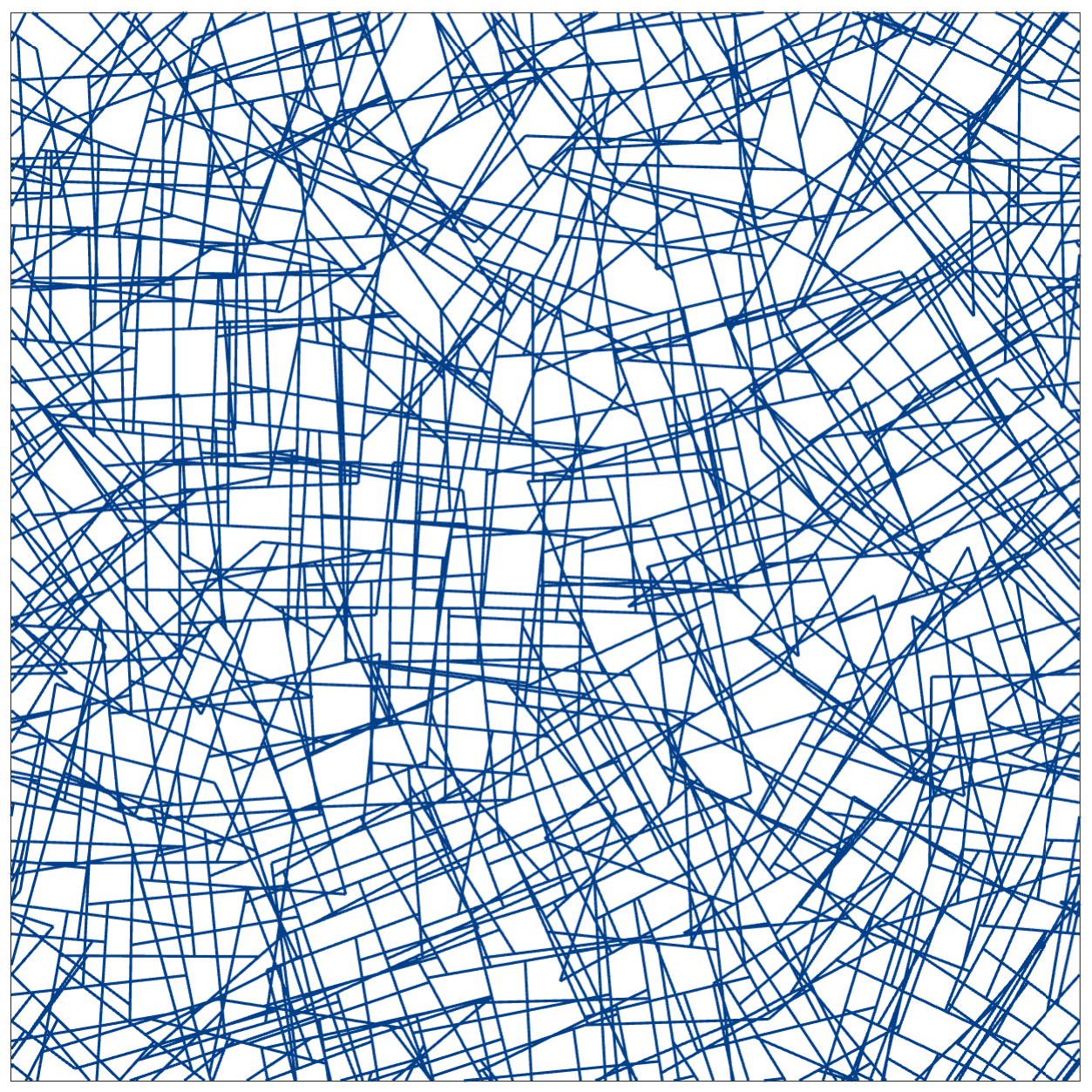}
         \put(5, 5){\colorbox{white}{\textbf{DOM type2}}}
         \end{overpic}
     \end{subfigure}
     \hfill
     \begin{subfigure}[b]{0.32\textwidth}
         \centering
         \begin{overpic}[width=1\textwidth]{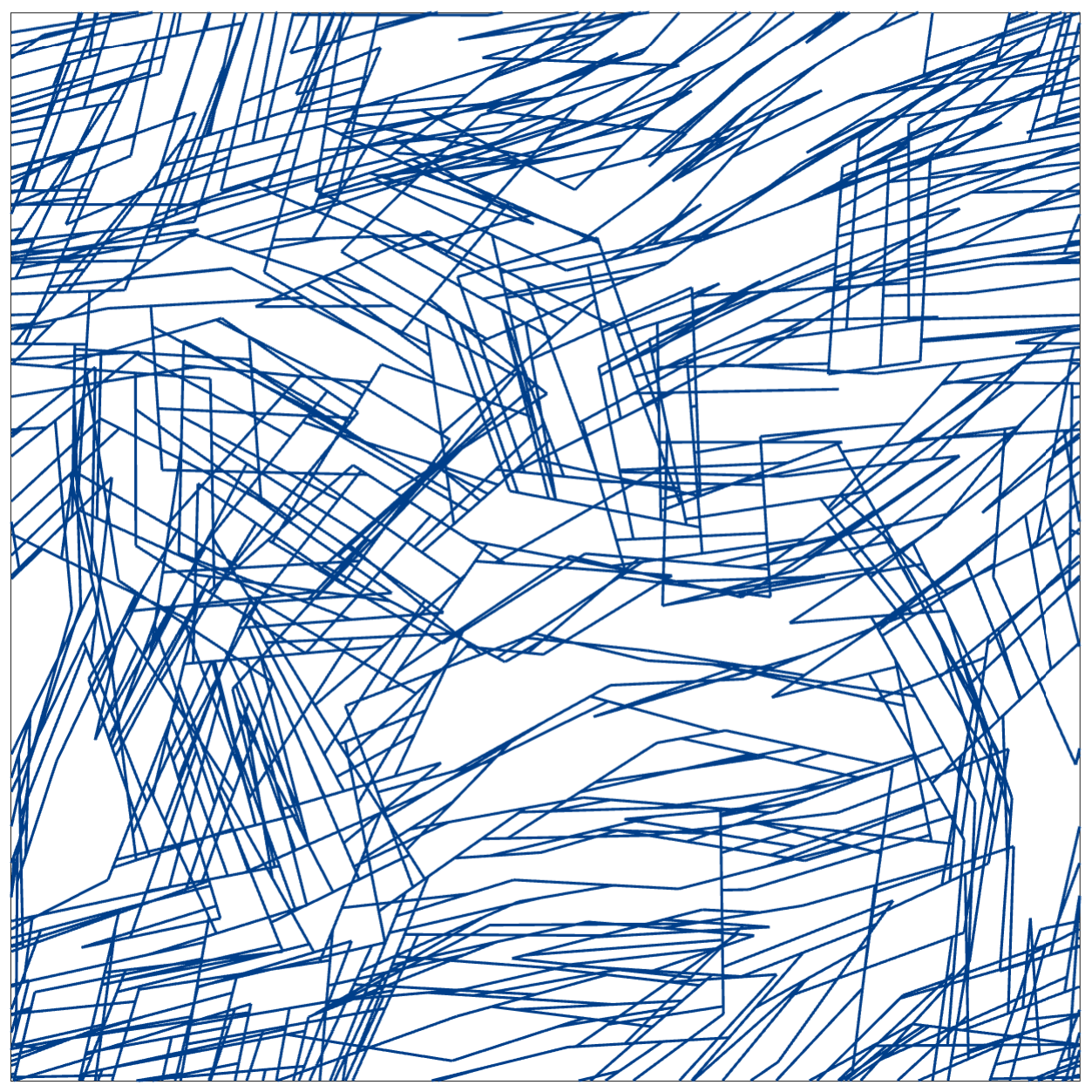}
         \put(5, 5){\colorbox{white}{\textbf{DOM type3}}}
         \end{overpic}
     \end{subfigure}
     \hfill
     \begin{subfigure}[b]{0.32\textwidth}
         \centering
         \begin{overpic}[width=1\textwidth]{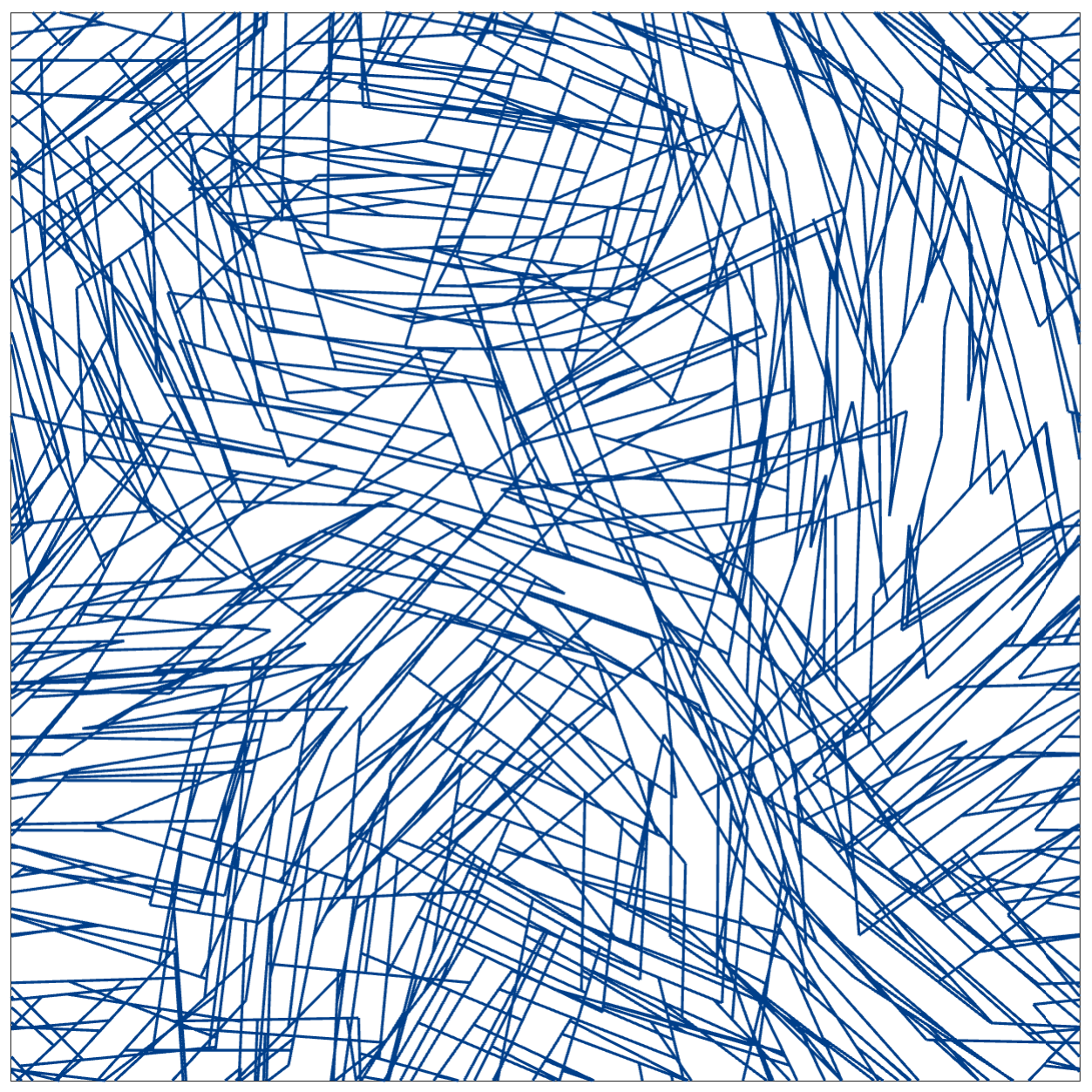}
         \put(5, 5){\colorbox{white}{\textbf{DOM type4}}}
         \end{overpic}
     \end{subfigure}
     \caption{Samples of the studied networks with $\rho=30$, $l=1$, and $L=5$. In the Mikado network, the red arrows indicate high density clusters.}
     \label{NT}
\end{figure}

We generated and compared three network classes: the unmodified Mikado network, the \textit{distribution-modified} (DM) network, where only spatial distribution is modified, and the \textit{distribution-orientation-modified} (DOM) network, where both spatial distribution and orientation are modified. For the DOM networks, we considered two orientations: $\Theta=0$ (which encourages fibres to co-align) and $\Theta=\pi/2$ (which encourages the fibres to be perpendicular). Networks with the parameters given in Table \ref{tab2} were studied, and examples of each type are shown in Figure \ref{NT}.

\begin{table}
\centering
\caption{Parameters used to generate the networks, the critical percolation threshold density, $\rho_c$, elastic exponent (from equation \eqref{Eper11}), $f$, and exponent at high density, $f_h$, for the six network classes considered (mean $\pm$ standard error).}
\begin{tabular}{cccccc}
\toprule
Network type&$P_r$&$\Theta$&$\rho_cl^2$&$f$ & $f_h$\\
\midrule
Mikado&N/A&N/A&$5.6373\pm0.0001$&$3.77\pm0.02$&$7.9\pm0.1$\\
DM&0.8&N/A&$4.9919\pm0.0002$&$3.56\pm0.01$&$7.7\pm0.1$\\
DOM 1&0&$\frac{\pi}{2}$&$4.954\pm0.001$&$3.70\pm0.01$&$8.4\pm0.1$\\
DOM 2&$0.8$&$\frac{\pi}{2}$&$4.1907\pm0.0001$&$3.53\pm0.01$&$8.0\pm0.1$\\
DOM 3&$0$&$0$&$9.43\pm0.01$&$3.92\pm0.02$&$9.6\pm0.1$\\
DOM 4&$0.8$&$0$&$7.690\pm0.001$&$3.68\pm0.02$&$9.0\pm0.1$\\
\bottomrule
\end{tabular}
\label{tab2}
\end{table}

\section{Results}
\subsection{Geometrical features}

We first studied percolation. All the subsequent analysis was carried out on percolated systems. Network percolation was investigated based on the finding that $\rho_{0.5}(L)$ (the density at which the percolation probability is 0.5 for finite $L$) converges to $\rho_c$ as $L\rightarrow\infty$ \cite{lij2009,ziff-92,ziff-2002}, since
\begin{equation}
 \rho_{0.5}(L)-\rho_c=-\frac{b_0}{a_1}L^{-1-1/\nu_c}+...~,
\label{PerConv}
\end{equation}
where $\rho_c$, $\nu_c$, $b_0$, and $a_1$ are the critical percolation density, the correlation-length exponent (a universal constant that is only affected by dimensionality, which equals $4/3$ for 2D systems \cite{lij2009,Stauffer_2018}), and some fitting constants, respectively. We  used a method similar to that developed by \citet{pike1974} to find  $\rho_{0.5}(L)$ (see Supplemental Material for further details).

The percolation thresholds are presented in Table \ref{tab2}. Our results for Mikado networks are consistent with the literature \cite{lij2009,mertens-2012}; however, as we modify the spatial dispersion, or make the fibres more likely to be perpendicular, $\rho_c$ decreases, and if we make the fibres more likely to be parallel, $\rho_c$ increases.

\begin{figure}
   \centering
     \begin{subfigure}[b]{0.49\textwidth}
         \centering
         \includegraphics[width=\textwidth]{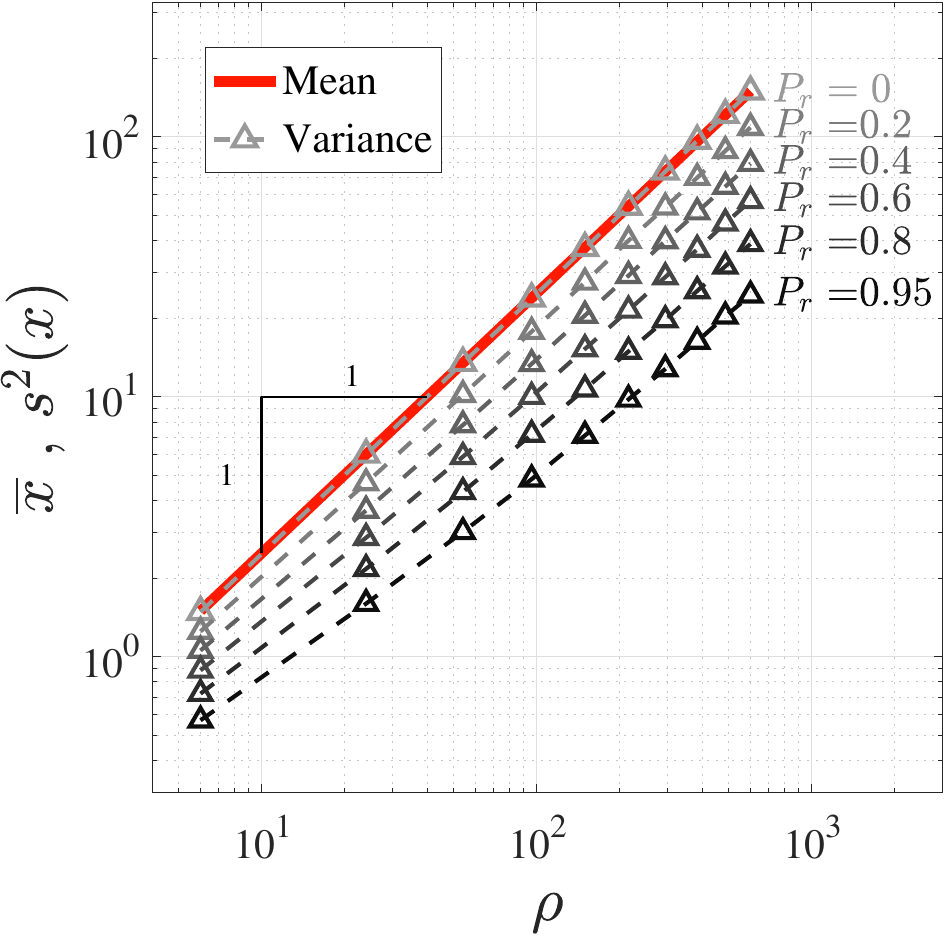}
         \caption{}
         \label{Dis}
     \end{subfigure}
     \hfill
     \begin{subfigure}[b]{0.49\textwidth}
         \centering
         \includegraphics[width=\textwidth]{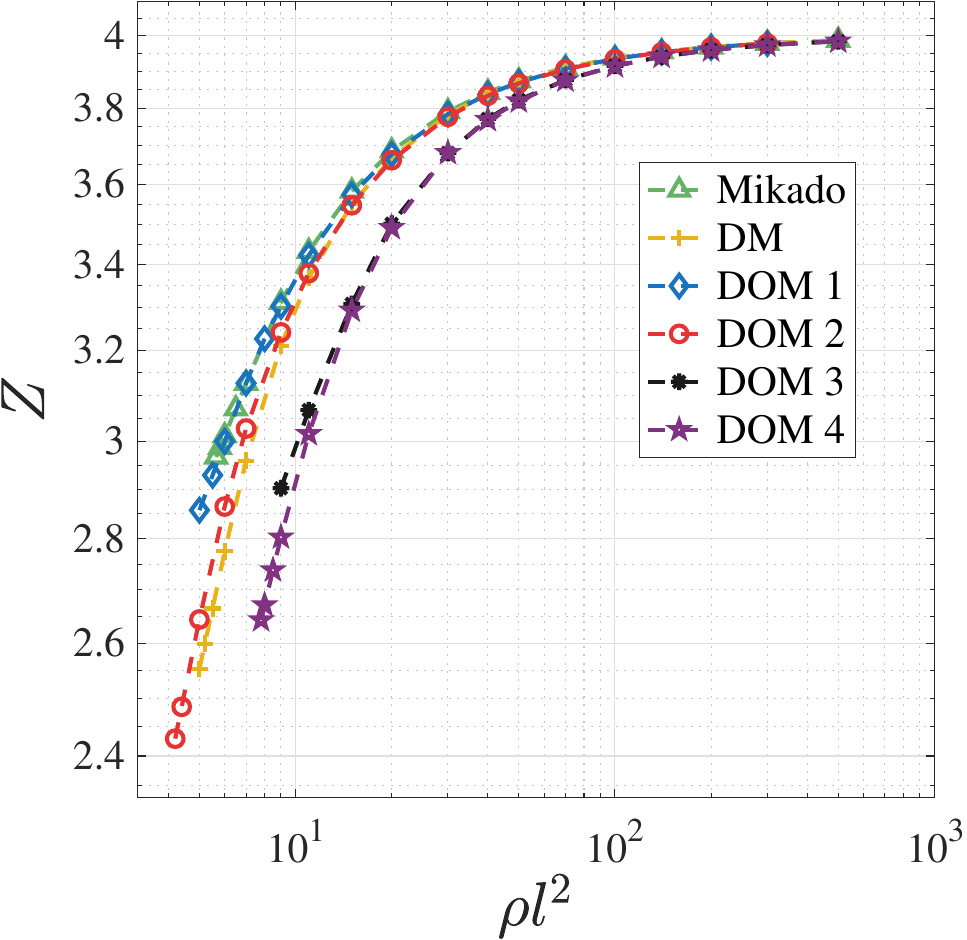}
         \caption{}
         \label{Z}
     \end{subfigure}
     \hfill \\
      \begin{subfigure}[b]{0.49\textwidth}
         \centering
         \includegraphics[width=\textwidth]{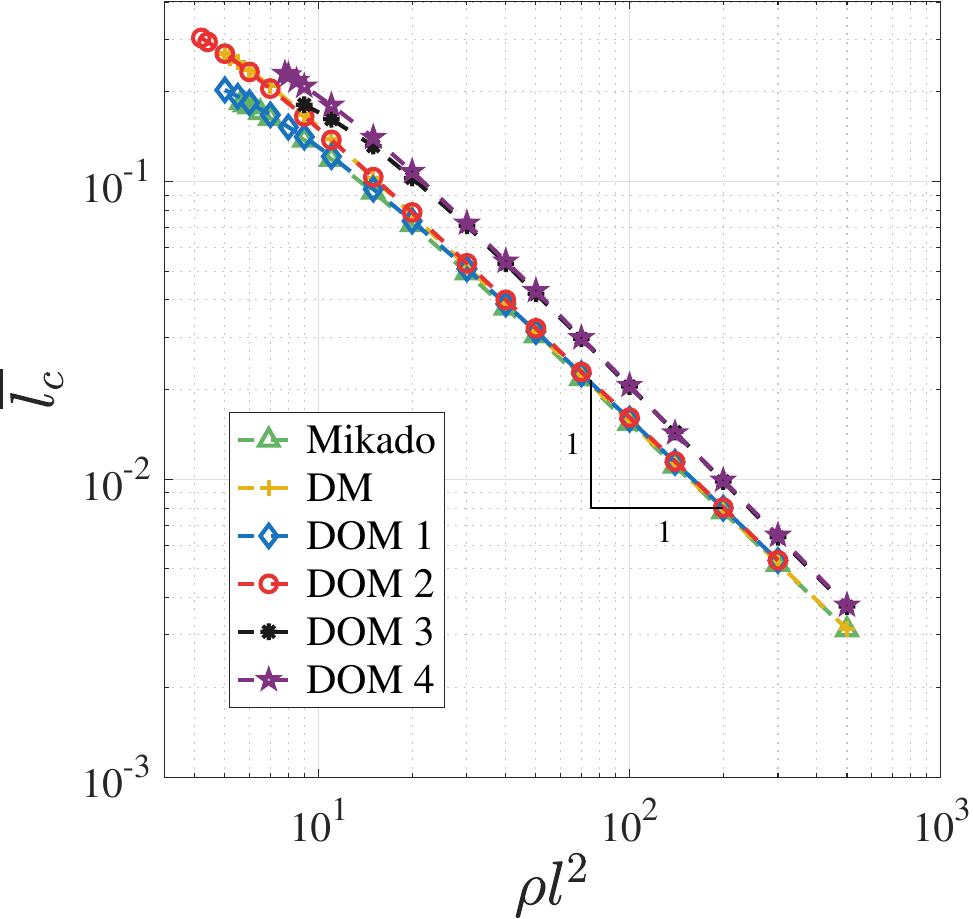}
         \caption{}
         \label{ml}
     \end{subfigure}
     \hfill
     \begin{subfigure}[b]{0.49\textwidth}
         \centering
         \includegraphics[width=\textwidth]{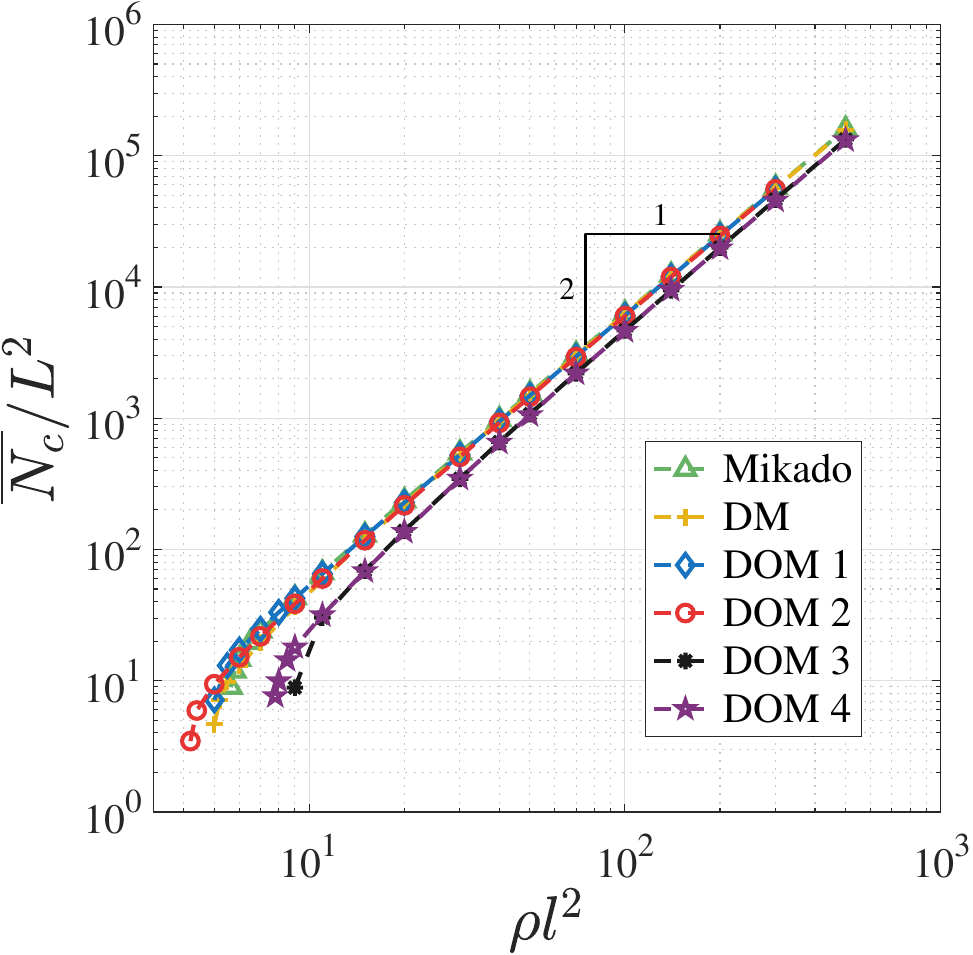}
         \caption{}
         \label{ne}
     \end{subfigure}
     \caption{Geometric features of the networks. (a) The mean, $\overline{x}$, and variance, $s^2$, of the number of fibres in 400 sub-boxes, as described in the text, for Mikado Networks and DM networks with various values of $P_r$. (b) The connectivity, $Z$, as a function of the dimensionless density, $\rho l^2$. (c) The mean segment length, $\overline{l_c}$. (d) The mean number of segments per unit area, $\overline{N_c}/L^2$. The standard errors for this data are smaller than the size of the markers in all plots.}
     \label{fig2}
\end{figure}

The second aspect that we investigated was how fibres are dispersed within the domain of study. To do so, we created networks of various densities with $L=10$, divided them into 400 smaller boxes, counted the number of fibre centres in each box, then calculated the mean and variance of this quantity. It is well known that, in Mikado networks, the number of fibre centres in each box follows a Poisson distribution \cite{sampson2009} (Figure \ref{Dis}). An increase in the value of $P_r$ results in a more homogeneous (lower variance) network (with fewer high density clusters, as indicated in the Mikado network in Figure \ref{NT}).

It has been shown that the connectivity of a network, $Z$, plays an important role in the network's behaviour \cite{Parzez2022}. As depicted in Figure \ref{Z}, for more homogeneous networks, the connectivity is slightly lower. The reason for this is fibre clustering (see Figure \ref{NT}) \cite{sampson2009}. As a result, fibres in the vicinity of these clusters have a relatively high chance of forming connections, and thus will have a higher connectivity. Moreover, for the networks whose fibres have a preference for being parallel ($\Theta=0$), the connectivity is considerably smaller. However, the difference between all network classes vanishes as the density increases.

Two other important geometrical factors are the mean length and density of fibre segments, as plotted in Figures \ref{ml} and \ref{ne}. Networks with $\Theta=0$ have longer fibre segments and lower segment density but gradually converge with the other networks as $\rho l^2\rightarrow\infty$. It will be shown that this affects the modulus of the networks substantially. In contrast, other network classes converge at lower densities, suggesting relatively similar mechanical responses.

\subsection{Mechanical behaviour}

Our investigation into the networks' mechanical properties is divided into the two cases of near percolation and high density \cite{Head_2003,Wilhelm_2003}. For low densities, the finite size scaling method \cite{Stauffer_2018,Wilhelm_2003} is used, and at high densities, a representative volume element (RVE) is considered \cite{Shahsavari_2013}.
In this study, the segments of the network are considered to be beams with second moment of area, $I$ \cite{logan2016}, with square cross-sections. The nodes are considered to be stiff cross-links (relative rotation is prohibited between beams connected at a node). The Timoshenko beam model \cite{logan2016} is used to model the beams and the direct stiffness method \cite{logan2016} is used to model infinitesimal deformation of the network. In all of our simulations, the length of each fibre is one, the thickness and density are varied as mentioned in each section, and the reported moduli are normalized on the fibres' Young's modulus.

It has been shown that the following scaling law can be used to describe the relationship between the mean Young's modulus, $\overline{E}$, and the size of the network at $\rho=\rho_c$, \cite{zabolitzky-1986}
\begin{equation}
\overline{E}=c_1 L^{-f/\nu_c},
\label{Eper11}
\end{equation}
where $c_1$ is a fitting constant. This is equivalent to $-\frac{\ln{\overline{E}}}{\ln{L}} = \frac{f}{\nu_c}-\frac{\ln{c_1}}{\ln{L}}$. By calculating $\overline{E}$ for various system sizes, and using a plot with $-\frac{\ln{\overline{E}}}{\ln{L}}$ and $\frac{1}{\ln{L}}$ as the axes, the intersection of the fitted curve with the $-\frac{\ln{\overline{E}}}{\ln{L}}$ axis is $\frac{f}{\nu_c}$.

To calculate the elastic modulus, a displacement in the $y$-direction was applied to each node on the top boundary, while the displacement in the $x$-direction and rotation were fixed. On the bottom boundary, zero displacement and rotation was imposed, and on the lateral boundaries, zero force was imposed. The number of realisations varied with system size from 1,600 for the largest to 102,600 for the smallest (see the Supplemental Material for further details).

The results are presented in Table \ref{tab2}. For the Mikado networks, the value of the elastic exponent was calculated as $f=3.73\pm0.02$. This is different from a previously reported value, $f=3.2\pm0.3$ \cite{Wilhelm_2003,Head_2003}, due to our use of a more accurate value for the percolation threshold. Our results show that the various network types considered here have different elastic exponents. This suggests that materials whose fibres interact during formation, thus modifying the structure of the network, have mechanical behaviour that scales differently with density to materials whose fibres do not interact. The value of the elastic exponent near percolation specifies how the material stiffens. The value of $f$ is especially important for materials whose fibre density is close to the percolation density, such as particulate gels. It has recently been discussed by \citet{Bantawa2023} that one possible reason for the difference in the observed exponent in colloidal gels is the fractal nature of the edges in the structure. They introduced an equation that relates the fractal dimensionality of the strains to the observed exponent: $f_{obs}=f/\nu_c(d-d_f)$, where $f_{obs}$, $\nu_c$, $d$, and $d_f$ are the observed exponent, correlation exponent, dimensionality, and fractal dimensionality of the gel chains, respectively. However, they considered the exponent $f$ to be constant for all of their systems. We have shown that $f$ changes when the network structure changes in the ways described above.

It has been shown that, at high density, the behaviour of fibre networks can be divided into two regimes: bending dominated and stretching dominated \cite{Shahsavari_2013, Head_2003,Wilhelm_2003} (we have defined the border of the bending dominated regime as being the density threshold below which more than $99\%$ of the total energy in the network is stored in bending). In the bending dominated regime, the bending stiffness of the fibre segments is much smaller than their stretching stiffness. In the stretching dominated regime, the network elastic moduli vary linearly with density and in the bending dominated regime, they vary proportionally to $(\rho l)^{f_h}(l_b/l)^2$, where $l_b=\sqrt{I/A}$, and $A$ is the fibre cross-sectional area \cite{Shahsavari_2013}.
 
\begin{figure}
   \centering
     \begin{subfigure}[b]{0.45\textwidth}
       \centering
         \includegraphics[width=\textwidth]{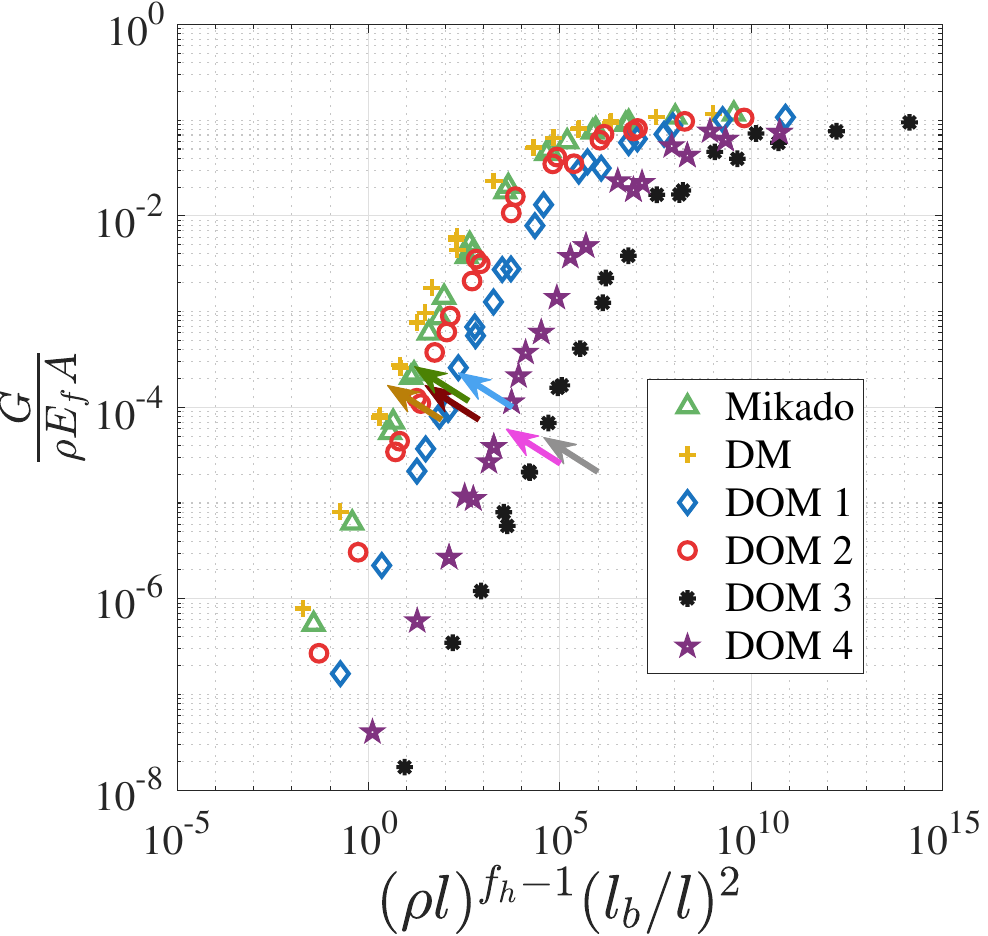}
         \caption{}
         \label{GHmaster}
     \end{subfigure}
     \hfill
     \begin{subfigure}[b]{0.45\textwidth}
       \centering
         \includegraphics[width=\textwidth]{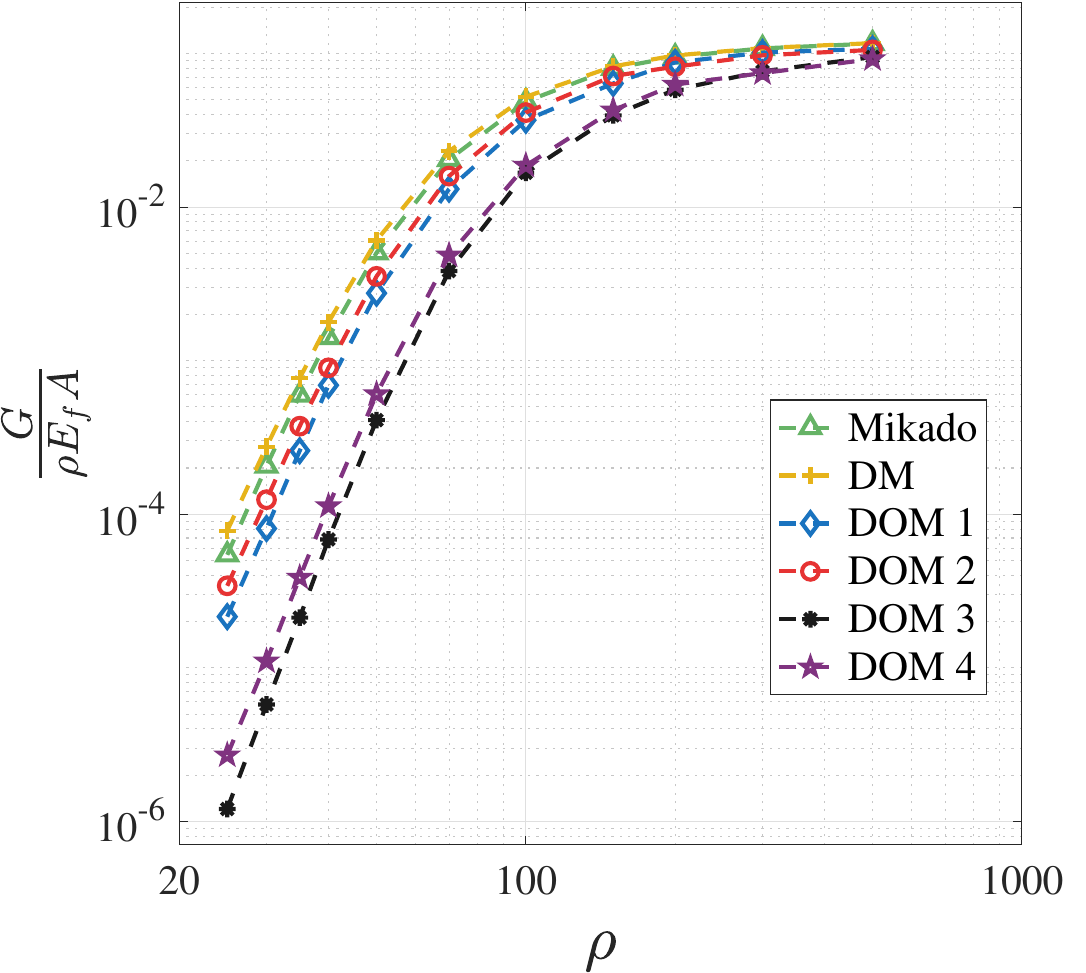}
         \caption{}
         \label{GH}
     \end{subfigure}
     \caption{(a) Master curves for the (non-dimensionalised) shear modulus. The values of $f_h$ are reported in Table \ref{tab2}. The arrows indicate the boundary of the bending dominated regime. (b) (Non-dimensionalised) shear modulus as a function of density for high density networks.}
     \label{highE}
\end{figure}

To determine the shear modulus, $G$, of the simulated networks, a periodic boundary condition with a far-field simple shear deformation was applied in a manner that is explained in depth by \citet{NGUYEN2012390}. In order to study $G$ for different network types, we simulated networks with fibre thicknesses ranging from $10^{-5}$ to $10^{-3}$ and densities ranging from 25 to 500; the results are presented in Figure \ref{highE}. In Figure \ref{GHmaster}, we used several different values of $l_b$ corresponding to fibre thicknesses in the stated range (by plotting $G/\rho E_fA$, where $E_f$ is the fibre Young's modulus, as a function of $(\rho l)^{f_h-1}(l_b/l)^2$, the data for each network type collapse onto a single curve called a master curve by some authors), whereas in Figure \ref{GH}, we used a fixed fibre thickness of $10^{-4}$.

The value of $G$ varies significantly between the six network types. It can vary up to 50 times in the bending-dominated regime (peptide hydrogels exist mainly in this regime). For DOM type 3 and 4 networks, $G$ is smaller by at least one order of magnitude in the bending dominated regime. This is due to their lower segment density, and higher segment length, and, for the same reason, the transition away from the bending dominated regime occurs later. This clearly shows how influential network structure is on the mechanical behaviour of fibrous materials.

For the Mikado network, $f_h$ was found to be 7.9, as reported in the literature \cite{Shahsavari_2013,Shahsavari_2013_2}. The other values are presented in Table \ref{tab2}. Increasing the spatial dispersion of fibres slightly decreases the value of $f_h$. Furthermore, noticeable changes can be observed for networks with preferred parallel orientations. All of the geometrical features discussed in Figure \ref{fig2} impact the network mechanical moduli to some degree, but disentangling every specific relationship between structure and mechanics is not straightforward. It has been shown that increasing the connectivity affects the elastic moduli \cite{Parzez2022}, but the changes in connectivity that we observe here (Figure \ref{Z}) are very small compared to those studied by \citet{Parzez2022}. For DOM type 3 and 4 networks, the slope of the mean segment length and the mean segment density are steeper; which, in the bending-dominated regime, causes the value of $f_h$ to be higher.

\begin{figure}
   \centering
   \begin{subfigure}[b]{0.3\textwidth}
         \centering
         \includegraphics[width=\textwidth]{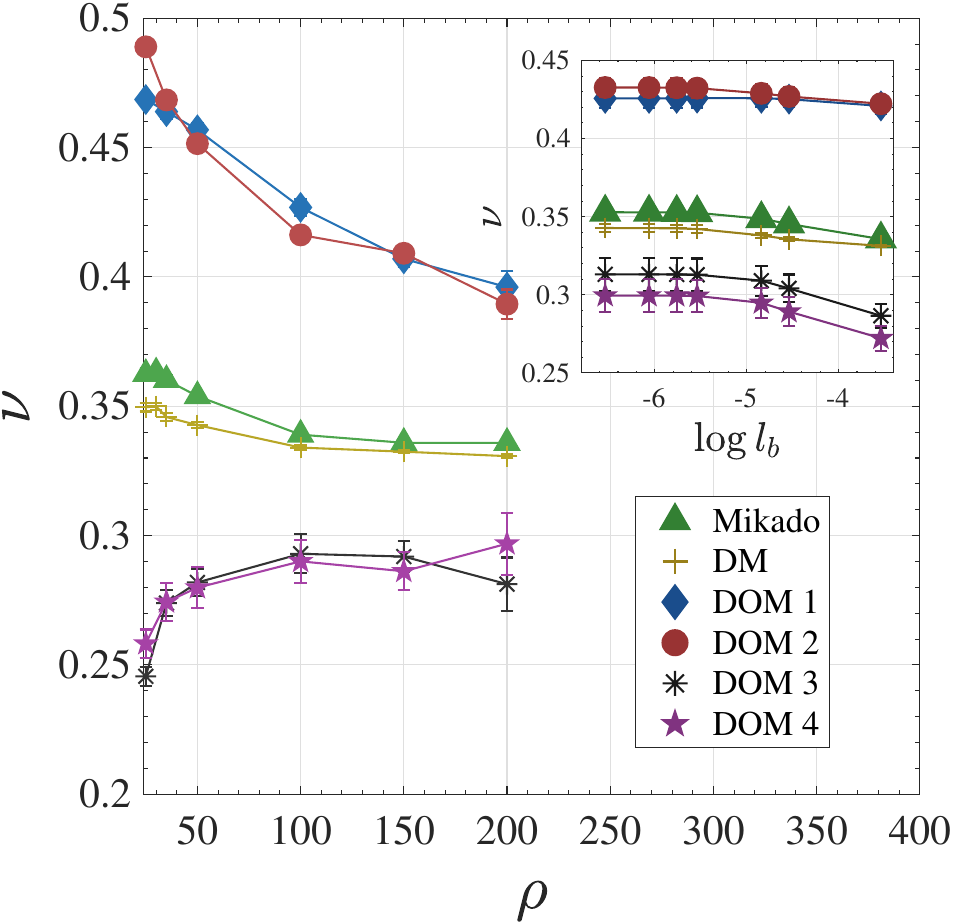}
         \caption{}
         \label{nu}
     \end{subfigure}
     \hfill
     \begin{subfigure}[b]{0.3\textwidth}
         \centering
         \includegraphics[width=\textwidth]{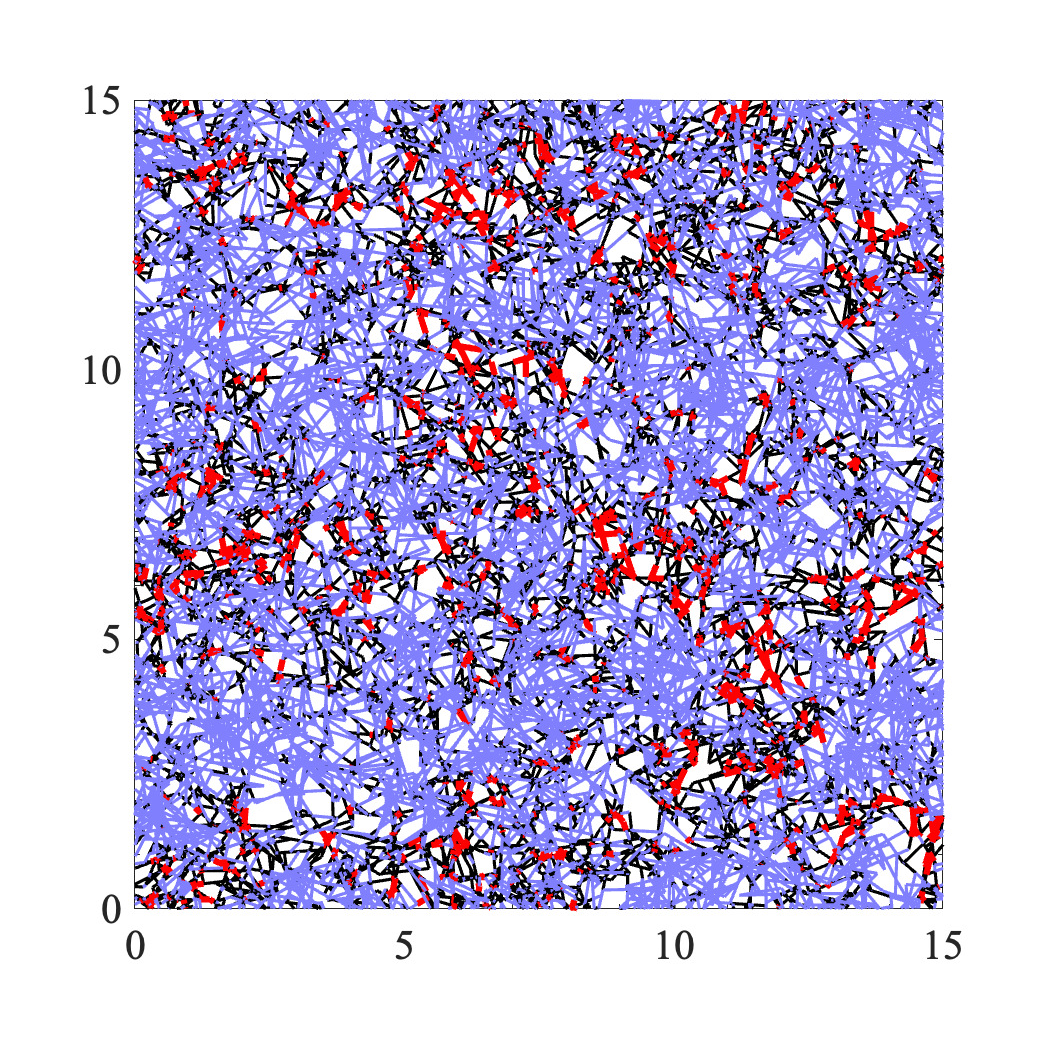}
         \caption{}
         \label{snaplow}
     \end{subfigure}
     \hfill
     \begin{subfigure}[b]{0.3\textwidth}
         \centering
         \includegraphics[width=\textwidth]{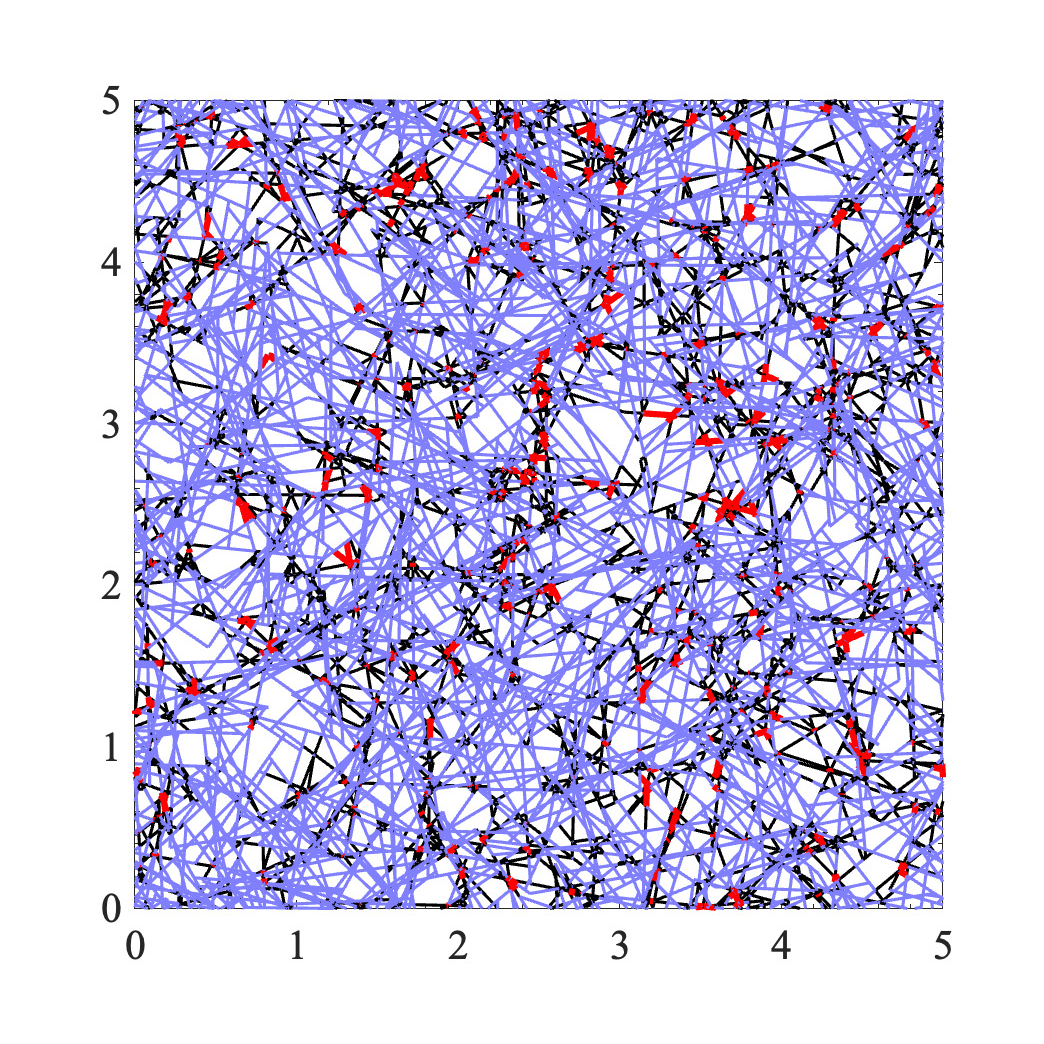}
         \caption{}
         \label{snaphigh}
     \end{subfigure}
     \hfill \\
      \begin{subfigure}[b]{0.3\textwidth}
         \centering
         \includegraphics[width=\textwidth]{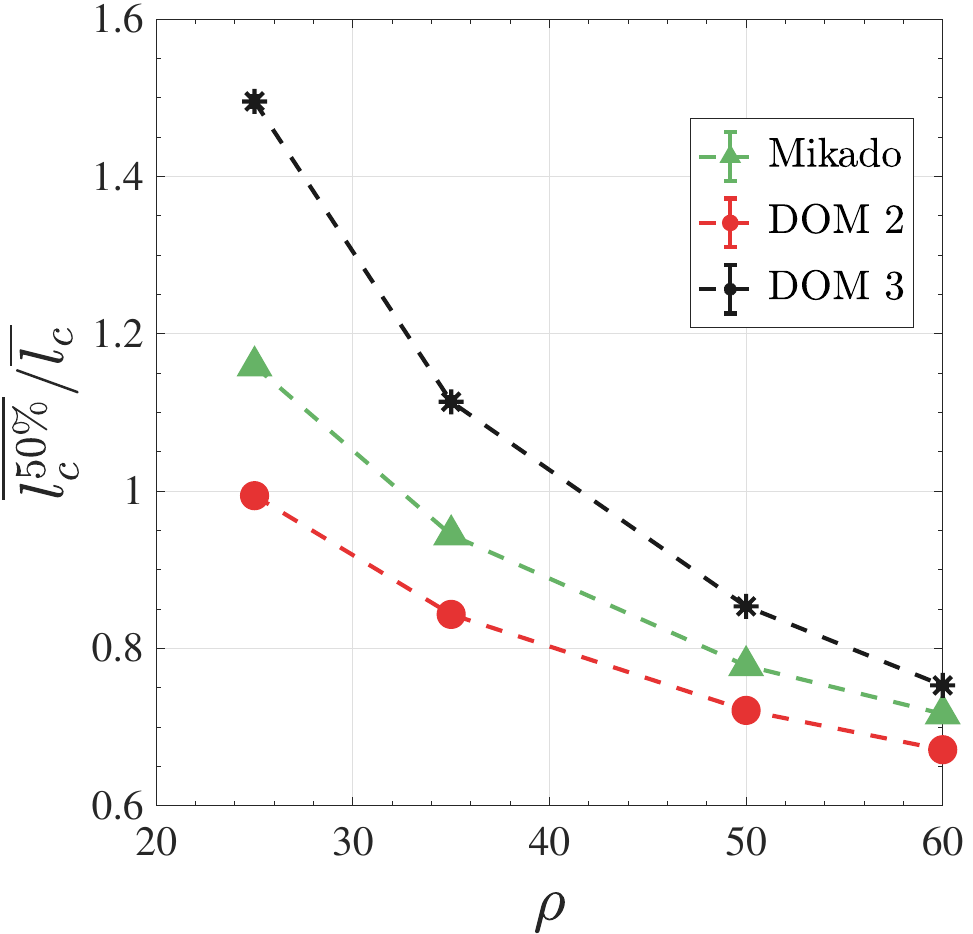}
         \caption{}
         \label{nulen}
     \end{subfigure}
     \hfill 
     \begin{subfigure}[b]{0.3\textwidth}
         \centering
         \includegraphics[width=\textwidth]{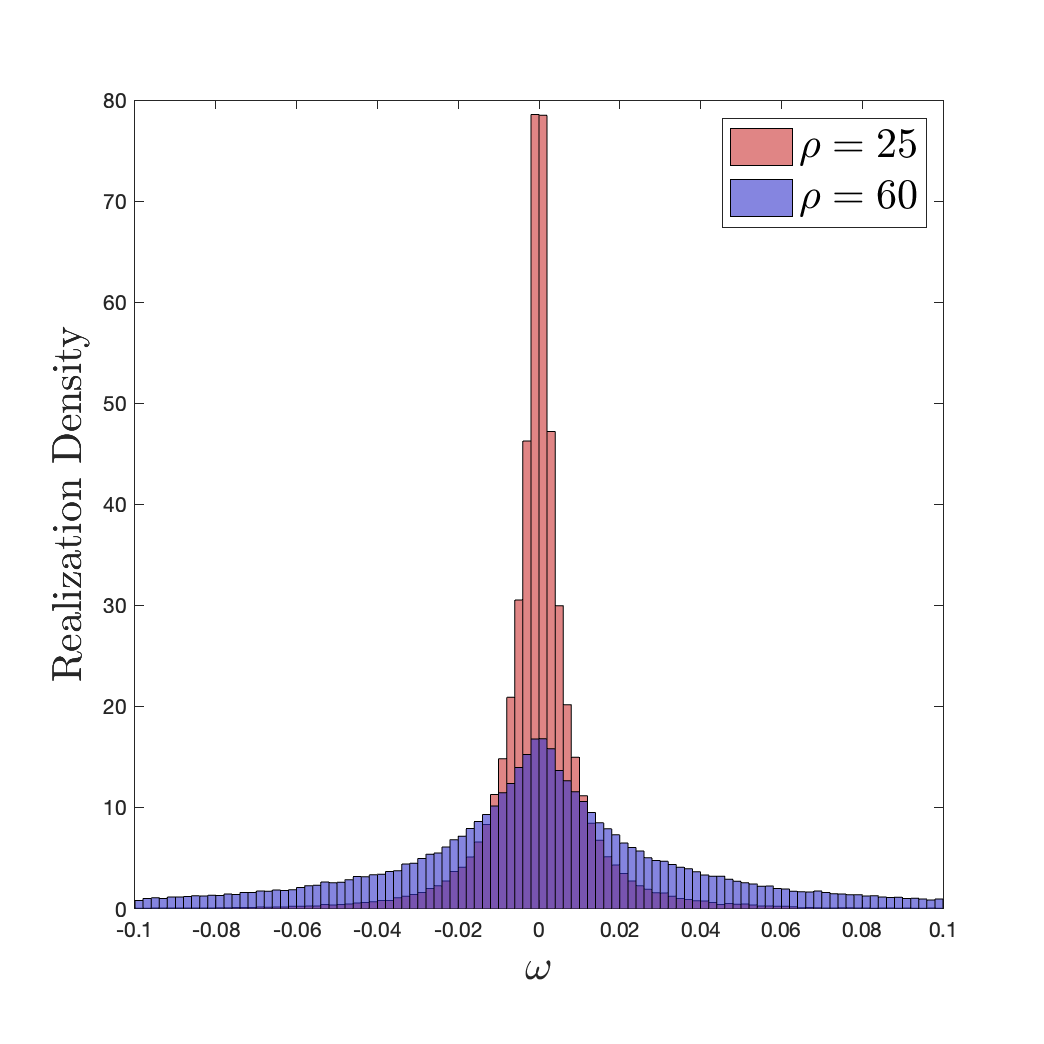}
         \caption{}
         \label{rothis}
     \end{subfigure}
     \hfill 
     \begin{subfigure}[b]{0.3\textwidth}
         \centering
         \includegraphics[width=\textwidth]{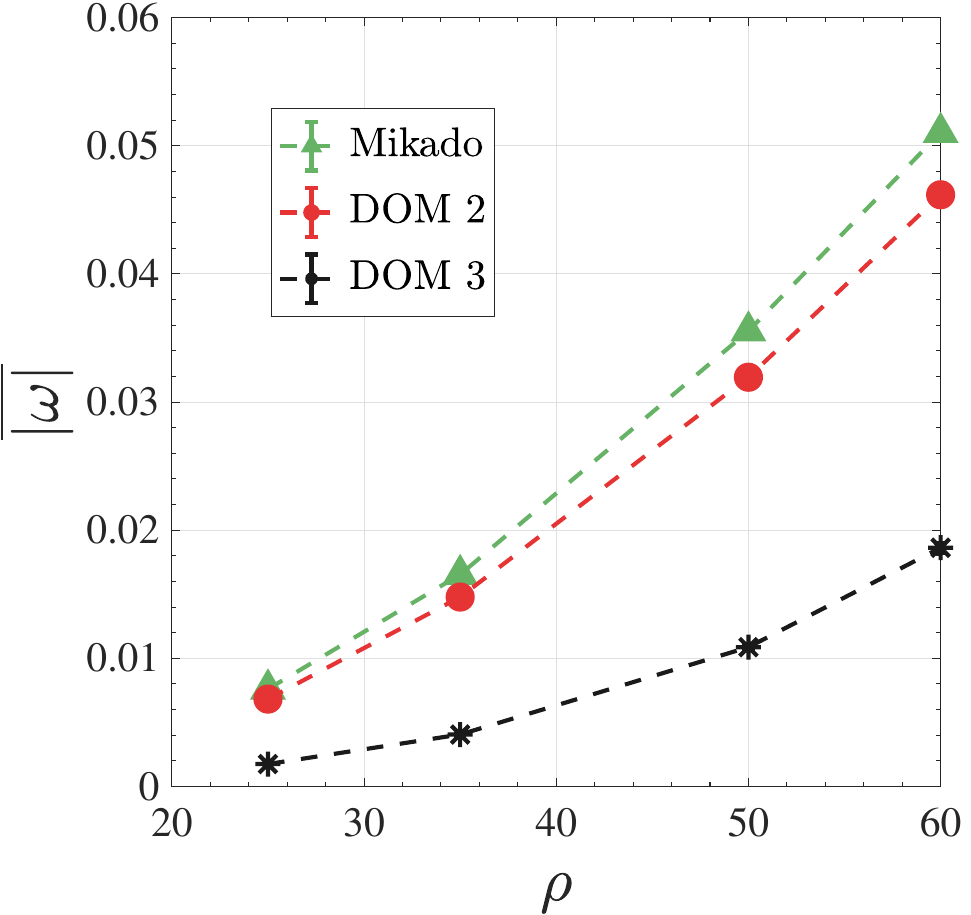}
         \caption{}
         \label{averot}
     \end{subfigure}
     \caption{(a) The dependence of Poisson's ratio on density and $l_b$. For the main panel, $l_b=2.9\mathrm{e}{-5}$. In the inset panel, $\rho=75$.(b) A snapshot of the Mikado network with density 25 and $L=15$. The red segments represent the smaller portion of the network that carries 50\% of the total energy of the system and the blue segments represent the largest possible portion of the network that only carries 5\% of the total energy. (c) A snapshot of the network with $\rho=60$ and $L=5$. (d) The average length of the segments that carry 50\% of the total energy, $\overline{l_c^{50\%}}$, divided by average segment length $\overline{l_c}$. (e) Histogram of node rotation, $\omega$, at two different densities at a strain of $1/1000$. (f) The average absolute value of rotation versus density.}
     \label{fig4}
\end{figure}

We also investigated the Poisson's ratio, $\nu$, of our six network types. To do so, we applied boundary conditions that impose a pure longitudinal deformation, and calculated the total forces in the directions parallel and perpendicular to the direction of stretch; Poisson's ratio is then calculated using $\nu=F_{\perp}/F_{\parallel}$, where $F_{\perp}$ and $F_{\parallel}$ are the total perpendicular and parallel forces at the boundary interfaces (see Supplemental Material for a derivation). We display the effects of changing the density, $\rho$, and $l_b$, in Figure \ref{nu}. There is a noticeable difference in the value of $\nu$ between the different networks, which is more pronounced when the relative orientations of the fibres change. 

The Poisson's ratio in the bending dominated regime is not constant and changes with increasing density. This is contrary to previous papers that reported $\nu$ to be constant  \cite{Head_2003,Shahsavari_2013_2}. The effect on $\nu$ of increasing density diminishes into the stretching dominated regime (as far as we have made observations). The effect of $l_b$, however, is different; in the bending dominated regime, increasing it has no effect, but as we transition into the stretching dominated regime, increasing it reduces the value of $\nu$ as far as we have made observations. 

As the density increases, how the network deforms changes even though we are still in the bending dominated regime -- this can be seen by examining how energy is stored in the network and how the nodes rotate (Figures \ref{snaplow} and \ref{snaphigh}). At smaller densities, the low density areas between clusters absorb energy (undergo deformation) and the clusters, where the density of segments is high and the average lengths are shorter, do not deform significantly. This means that these dense clusters move (rotate/translate) almost rigidly (with a low average contribution to absorbing energy). It can be observed, by looking at Figure \ref{nulen}, that, at low density, the average length of the segments that absorb most of the energy is higher than the average segment size of the network. As the density increases, this value decreases, and more fibres within the clusters contribute to absorbing energy. Furthermore, as the density increases, more segments deform (due to node rotation), as can be seen from Figure \ref{rothis} (note that the rotation of all nodes is zero for an affine deformation). The average absolute rotation is also plotted in Figure \ref{averot}, which shows that the same trend is followed by all network classes. Since the deformation of the microstructure changes as we increase the density, we may expect a change in the Poisson’s ratio (which is a direct result of microstructural movement). The low density deformation is greatly affected by the microstructure and the Poisson’s ratios are noticeably different for different network classes as a result. At higher densities, however, the Poisson's ratio values converge since, at higher densities, the interiors of the clusters also deform and, therefore, the effects of microstructure decrease gradually.

\subsection{Experiments}

To compare the results of the model networks qualitatively with real SAPHs, we carried out shear rheometry tests on three $\beta$-sheet forming peptides: KFEFEFKFK\textbf{F} (F), KFEFEFKFK\textbf{K} (K), and KFEFEFKFK\textbf{FK} (FK) (K: lysine; F: phenylalanine; E: glutamic acid). These peptides are based on a design first developed by \citet{zhang-1993,zhang-1999} (see the Supplemental Material for details). Figure \ref{tem}a-c shows transmission electron microscopy (TEM) images obtained for 10$\times$ diluted hydrogels alongside the chemical structure of the SAPHs. These images confirm the formation in all three systems of semi-rigid peptide fibre networks with the thinnest fibre diameter ranging from 3 to 4 nm. Considering the processing required, TEM images cannot give us completely accurate information about the \textit{in situ} structure of the networks directly. To obtain these images, we dilute the original hydrogel 10 times and then dry it on a grid. Due to this processing, some structural information is lost. However, we can conclude the following important points from the TEM images: 1) the fibres do not aggregate -- they are the same size in each network, which rules out mechanical effects that might arise from aggregation; 2) almost all of the peptides form fibres and they connect to each other (if this was not the case, a direct relation between density and peptide concentration would not exist); 3) the fibres form in approximately straight lines.

\begin{figure}
   \centering
     \begin{subfigure}[b]{0.49\textwidth}
         \centering
         \begin{overpic}[width=1\textwidth]{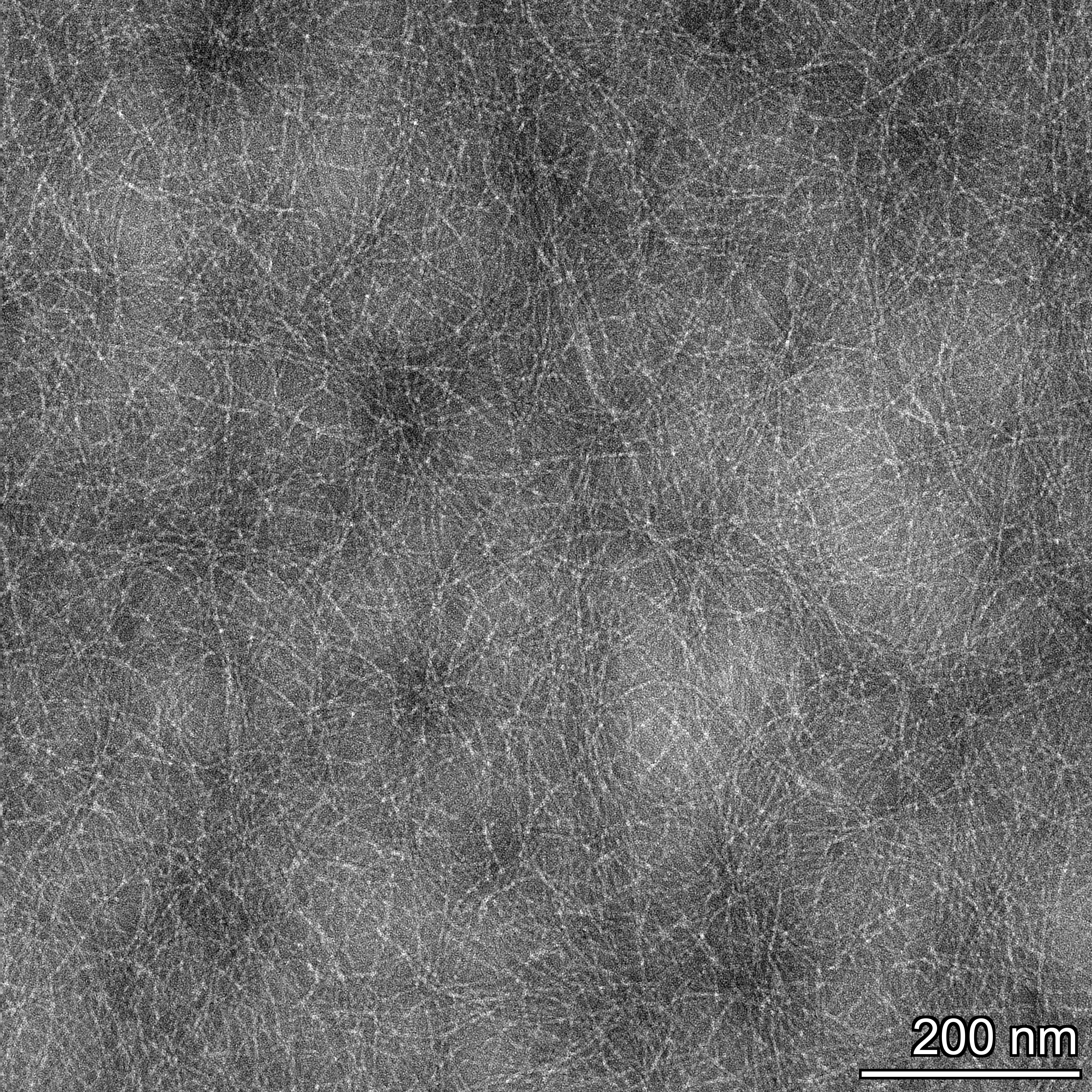}
         \put(5,70){\includegraphics[width=0.8\textwidth]{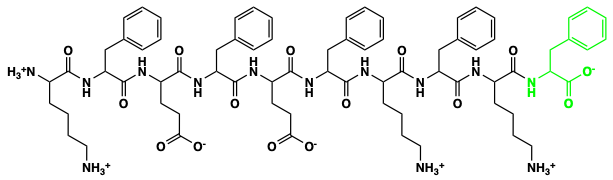}}
         \put(5, 5){\colorbox{white}{\textcolor{green}{\textbf{F}}}}
         \end{overpic}
         \caption{}
     \end{subfigure}
     \hfill
     \begin{subfigure}[b]{0.49\textwidth}
         \begin{overpic}[width=1\textwidth]{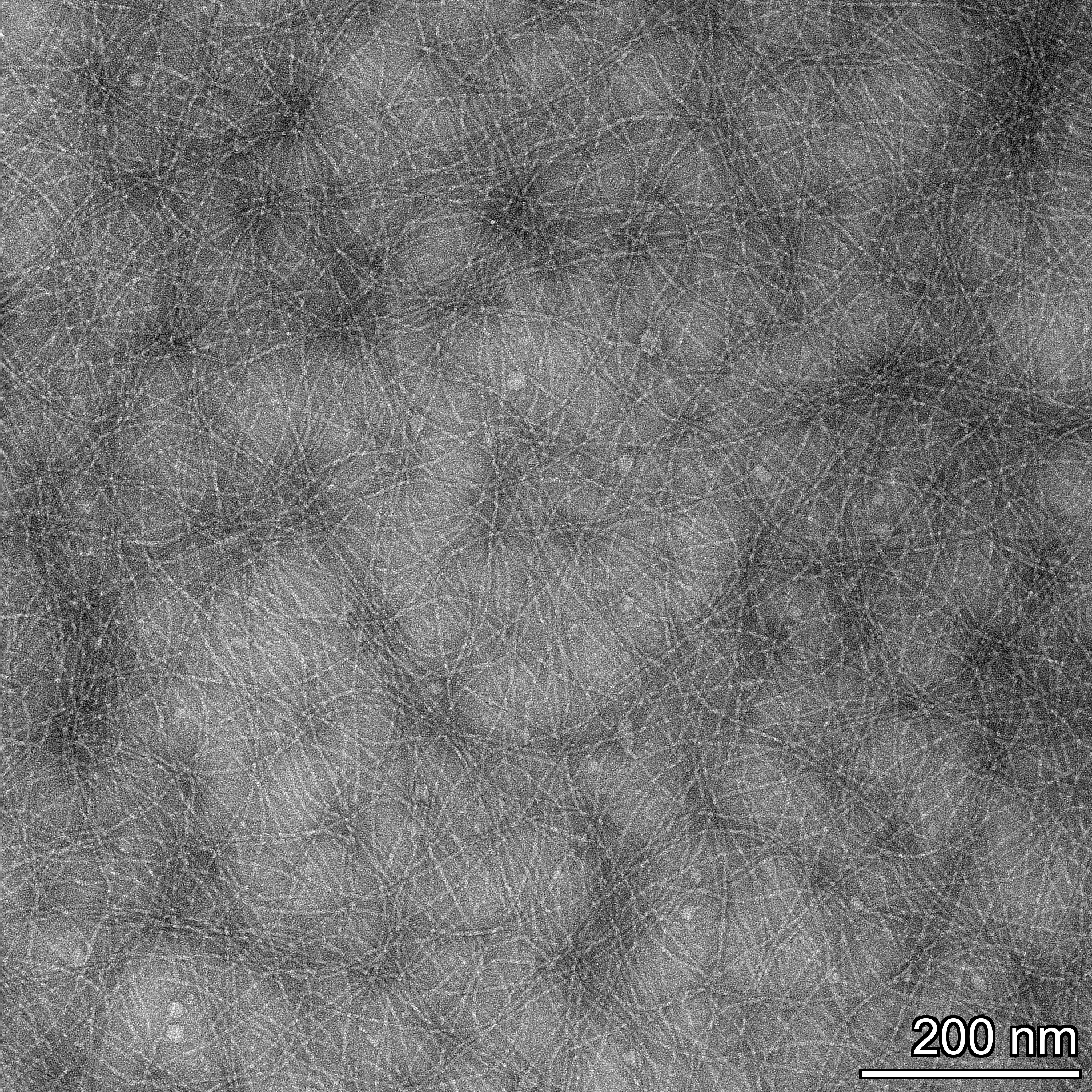}
         \put(5,70){\includegraphics[width=0.8\textwidth]{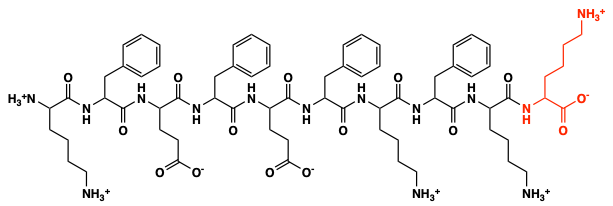}}
         \put(5, 5){\colorbox{white}{\textcolor{red}{\textbf{K}}}}
         \end{overpic}
         \caption{}
     \end{subfigure}
     \hfill \\
     \begin{subfigure}[b]{0.49\textwidth}
         \centering
         \begin{overpic}[width=1\textwidth]{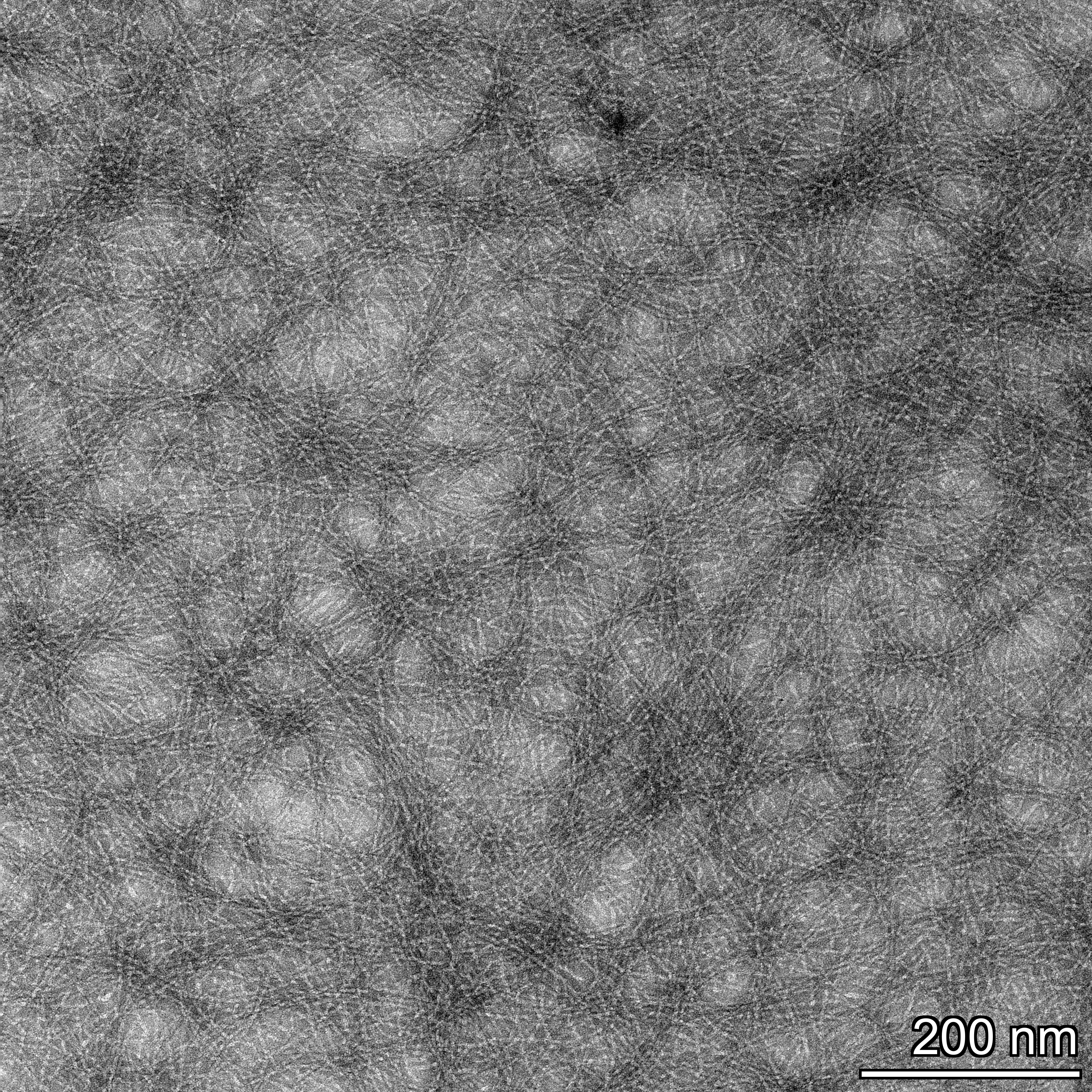}
         \put(5,70){\includegraphics[width=0.8\textwidth]{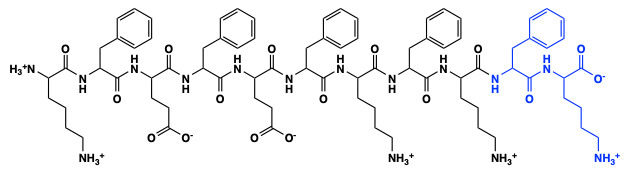}}
         \put(5, 5){\colorbox{white}{\textcolor{blue}{\textbf{FK}}}}
         \end{overpic}
         \caption{}
     \end{subfigure}
     \hfill
     \begin{subfigure}[b]{0.49\textwidth}
         \centering
         \begin{overpic}[width=1\textwidth]{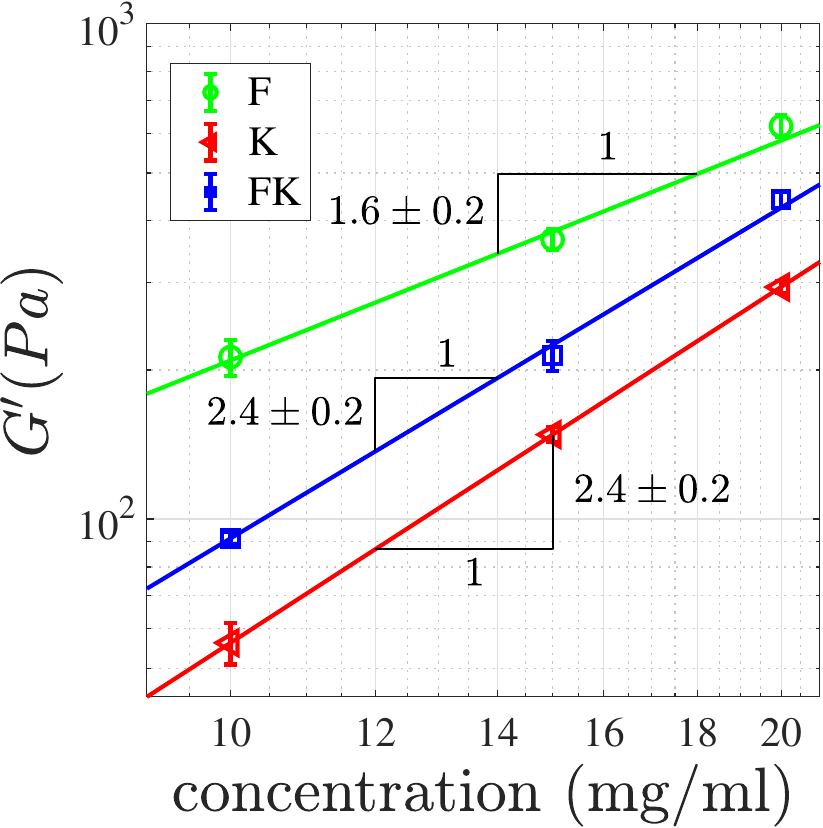}        
         \end{overpic}
         \caption{}
         \label{exp}
     \end{subfigure}
     \caption{(a)-(c) TEM images of three SAPHs with 10$\times$ dilution in 57000$\times$ magnification, and their chemical structure (inset). (d) Rheology test results for the SAPHs; $G^\prime$, the storage modulus, versus peptide concentration (which is a proxy for $\rho l$ in the models due to the one-to-one relationship between the mass and length of a peptide fibre).}
     \label{tem}
\end{figure}

\citet{Wychowaniec-2020} have shown that peptides with hydrophobic end-groups (F) have more hydrophobic edge interactions and are thought, therefore, to form more crosslinks. These networks are thus more similar to the perpendicular network model described above.
On the other hand, peptides with hydrophilic (K) end-groups were shown to make fewer crosslinks due to edge electrostatic repulsion and, therefore, will be more similar to the parallel network model. In Figure \ref{tem}d, the storage moduli, $G^\prime$, obtained for these three hydrogels are presented. For FK and K hydrogels, similar exponents (2.4 $\pm$ 0.2) were obtained. Both these systems have fibres with hydrophilic edges due to the presence of the terminal K. FK was found to have a higher $G^\prime$ compared to K which can be attributed to the presence of the additional phenylalanine group before the terminal K. This additional F increases the overall hydrophobicity of the peptides and the intermolecular forces (H-bond number) between peptides, resulting in more rigid fibres with higher moduli, without affecting the exponent. For the F hydrogels, significantly higher values of $G^\prime$ were observed (due to higher crosslinking levels) with a smaller exponent, as predicted by the models.

It has been observed that models with different dimensionality have qualitatively similar behaviour \cite{Arzash2021} but different exponents (the exponents of Mikado networks in 2D and 3D are 3 and 8, respectively \cite{Picu_book}). Despite the difference in the actual exponent observed in our model and in our experiments, the trends in the behaviour of the exponents are similar, and could arise from a similar phenomenon (the difference in the values originates from the difference in dimensionality, which is expected). For example, the attraction and repulsion forces between fibres would affect the structure of the network, and locally rearrange it, regardless of whether the system is 2D or 3D.

\section{Conclusion}

In conclusion, to investigate the effects of network structure on material properties, new networks based on Mikado networks were created and studied at low (near percolation) and high densities. Their behaviours were shown to be affected by localised modifications, which represent the existence of local interactions between the fibres when the network is forming. We found that increasing spatial dispersion, and modifying orientation such that fibres have a greater tendency to intersect, decreases the percolation threshold and elastic exponent near the percolation threshold. At high densities, in the non-affine regime, the trends in the model networks' shear moduli agreed with experiments we conducted on real SAPHs. Finally, we showed that the Poisson's ratio depends on both the deformation regime and the density. The reason for this lies in how the networks absorb energy at different densities.


\bibliography{refs}

\end{document}


\maketitle

\section{Percolation and Elastic Moduli Exponent Calculations}

\begin{figure} 
\centering
\includegraphics[width=0.6\textwidth]{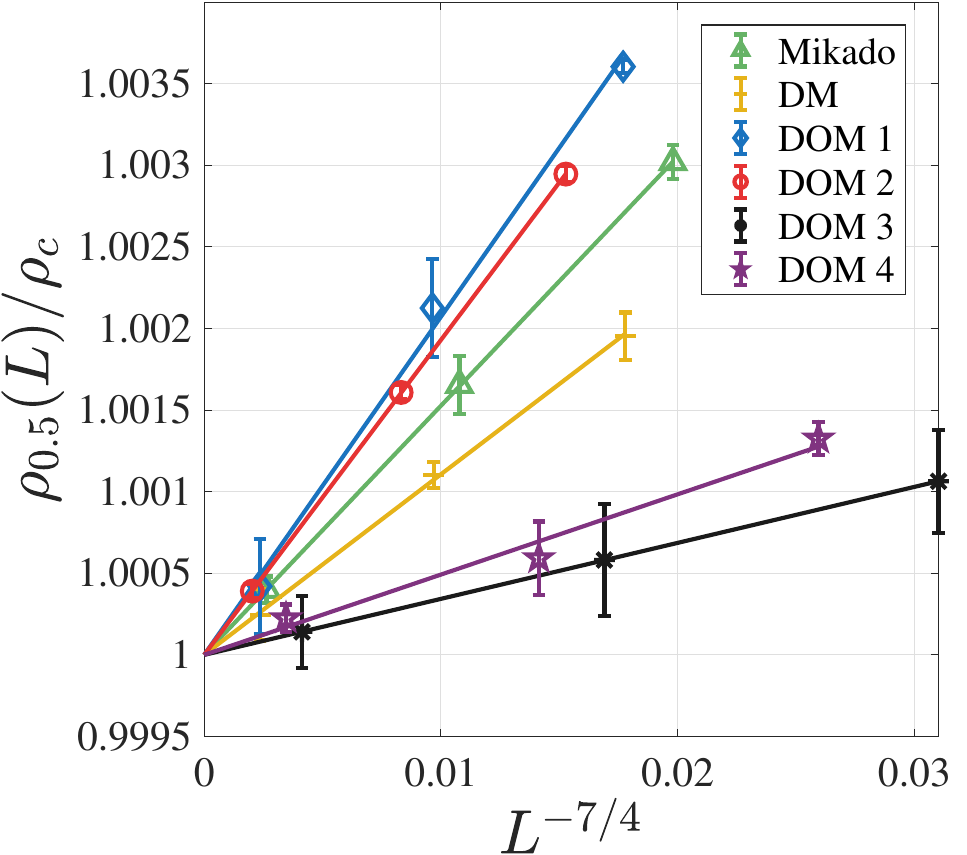}
\captionsetup{justification=centering}
\caption{Convergence of $\rho_{0.5}$ to $\rho_c$: as $L\rightarrow\infty$, $L^{-7/4}\rightarrow 0$ and $\rho_{0.5}/\rho_c\rightarrow 1$. The exponent $-7/4$ comes from equation \eqref{rho0.5}, since $\nu_c=4/3$ for two dimensional networks [6, 8].}
\label{per}
\end{figure}

The percolation probability, for a system with given side length, $L$, with density, $\rho$, near the percolation threshold, is [6,24,25]
\begin{equation}
R(\rho,L)=\frac{b_0}{L}+0.5+a_1y+a_2y^3+...,
\label{PerPro}
\end{equation}
where $y=(\rho-\rho_c)L^{1/\nu_c}\ll 1$, and $\rho_c$, $\nu_c$, $b_0$, $a_1$, and $a_2$ are the critical percolation density, the correlation-length exponent, and some fitting constants, respectively. From this equation, we deduce that [6,24,25]
\begin{equation}
\rho_{0.5}(L)-\rho_{c}=-\frac{b_0}{a_1}L^{-1-1/\nu_c}+...,
\label{rho0.5}
\end{equation} 
where $\rho_{0.5}(L)$ is the density at which the probability of percolation is 0.5 for a system of size $L$. We have neglected higher order terms (which is reasonable when $\rho$ is close enough to $\rho_c$ that $(\rho-\rho_c)^{2n+1}L^{2n/\nu_c}$ is negligible  $\forall n\in\mathbb{N}$). Once $\rho_{0.5}(L)$ is known for several system sizes, it is possible to calculate the percolation density to a desired accuracy using this equation. To calculate $\rho_{0.5}(L)$, we started with a system with $L=1$ and added a number of fibres (we chose to use 500, 1000 and 5000 fibres to give three data points for each network type). We adjusted the length of the fibres, $l$, until the probability of percolation equaled $0.5$. To do this, we first roughly estimated the value of $l$ (by doing a small number of simulations using trial and error to find a value of $l$ that gave a percolation probability close to 0.5 with a relatively high margin of error), we then considered six values of $l$ around the roughly estimated value, making $10^5$ (when there were 5000 fibres) or $10^6$ (when there were 500 or 1000 fibres) system observations for each value of $l$, and then calculated the value of $l$ that corresponded to a percolation probability of 0.5 by interpolating between these values. Next, we scaled the whole system to make $l=1$ and found the corresponding values of $L$ and $\rho_{0.5}(L)$ [26]. The values of $\rho_{0.5}(L)/\rho_c$ for our different network types are plotted in Figure \ref{per}.

\begin{figure} 
\centering
\includegraphics[width=0.55\textwidth]{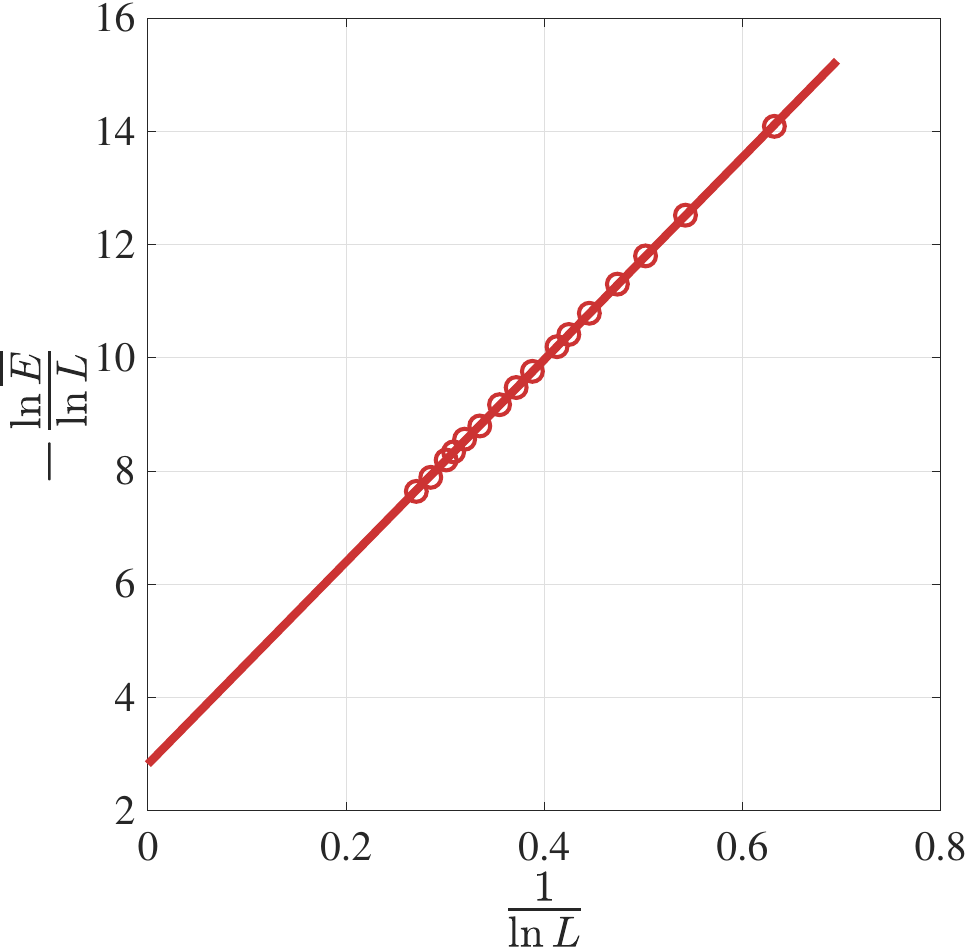}
\captionsetup{justification=centering}
\caption{The mean elasticity modulus, $\overline{E}$, for Mikado network with $l=1$ and $thickness = 0.001$ at various system sizes have been calculated and plotted. As mentioned in the text, they form a straight line in $-\frac{\ln{\overline{E}}}{\ln{L}}$ and $\frac{1}{\ln{L}}$ coordinates and the intercept is equal to $f/\nu_c$. The standard error for each data point is smaller than the markers.}
\label{LowE}
\end{figure}

It is known that the values of the elastic moduli of networks near percolation change exponentially with density; for example, for the Young's modulus, $E\propto(\rho-\rho_c)^f$. In order to have a good estimate of $E$ at each density, the system size must be much larger than the correlation length, $\xi$. Precisely at the percolation density, $\xi=\infty$, but as the density increases, the correlation length decays as $\xi\propto(\rho-\rho_c)^{-\nu_c}$ [8]. Therefore, the correlation length is very large near the percolation threshold, so directly investigating the behaviour of such networks is impractical. To overcome this issue, finite size scaling methods have been adopted. It has been shown that the following scaling law can be used to describe the relationship between the elastic moduli and the size of the network at $\rho=\rho_c$ [30],
\begin{equation}
\overline{E}=c_1 L^{-f/\nu_c},
\label{Eper11}
\end{equation}
where $\overline{E}$ and $c_1$ are the mean of $E$ and a constant, respectively. This is equivalent to
\begin{equation}
-\frac{\ln{\overline{E}}}{\ln{L}} = \frac{f}{\nu_c}-\frac{\ln{c_1}}{\ln{L}}.
\end{equation}
By calculating $\overline{E}$ for $\rho=\rho_c$, with various system sizes (only considering cases where the system is percolated), and using a plot with $-\frac{\ln{\overline{E}}}{\ln{L}}$ and $\frac{1}{\ln{L}}$ as the axes, the intersection of the fitted curve with the $-\frac{\ln{\overline{E}}}{\ln{L}}$ axis will be $f/\nu_c$. For the Mikado network class, the fitted line is plotted in Figure \ref{LowE}.

\section{Geometric Features Near the Percolation Threshold}

\begin{figure}
   \centering
     \begin{subfigure}[b]{0.31\textwidth}
         \centering
         \includegraphics[width=\textwidth]{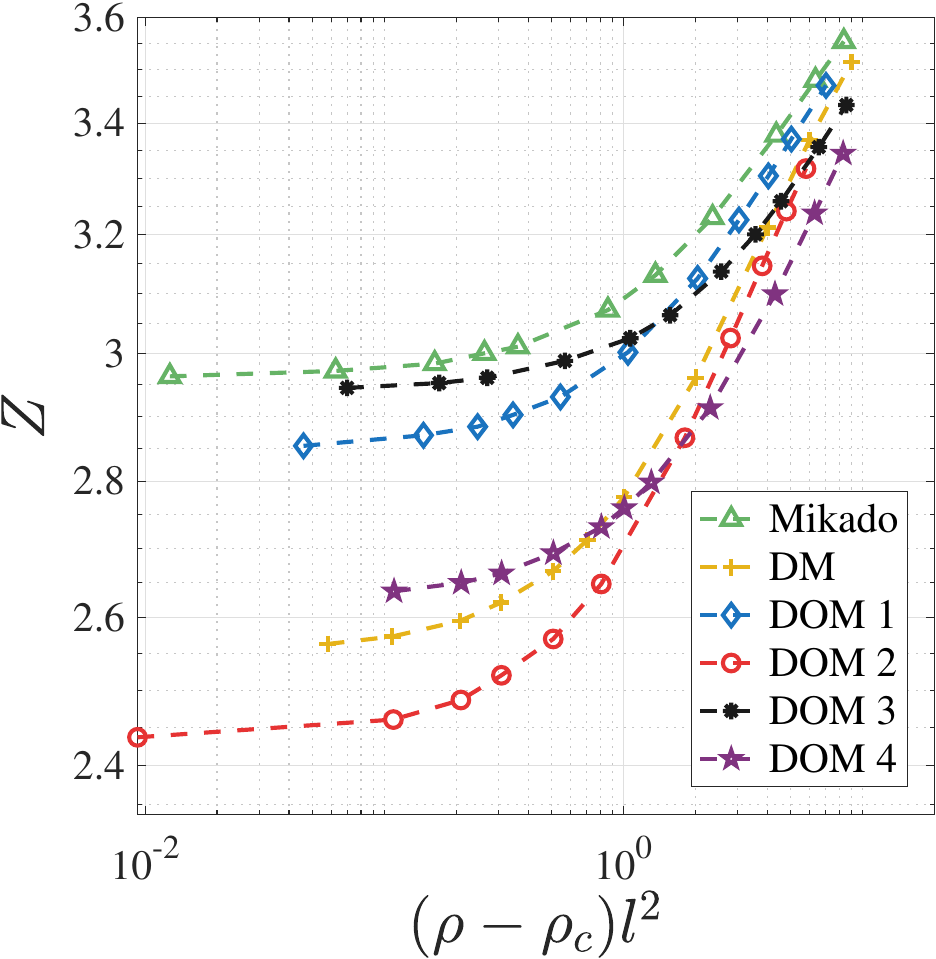}
         \caption{}
         \label{zper}
     \end{subfigure}
     \hfill
     \begin{subfigure}[b]{0.31\textwidth}
         \centering
         \includegraphics[width=\textwidth]{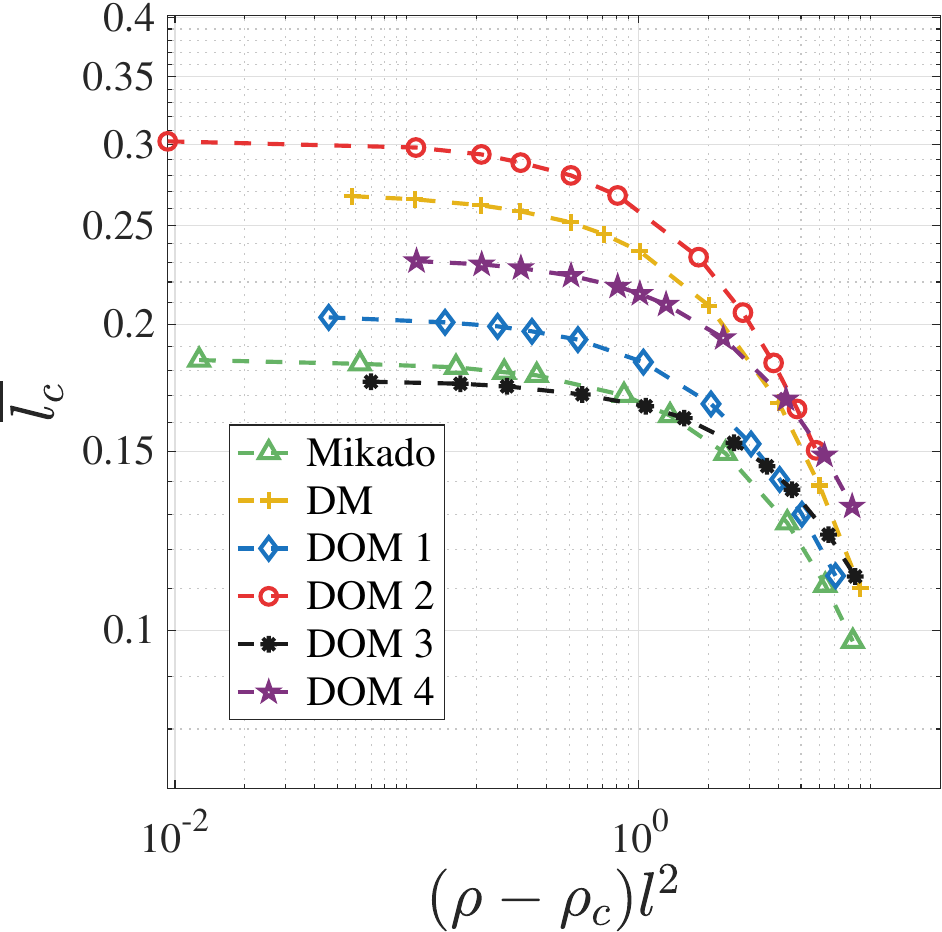}
         \caption{}
         \label{mlper}
     \end{subfigure}
     \hfill
      \begin{subfigure}[b]{0.31\textwidth}
         \centering
         \includegraphics[width=\textwidth]{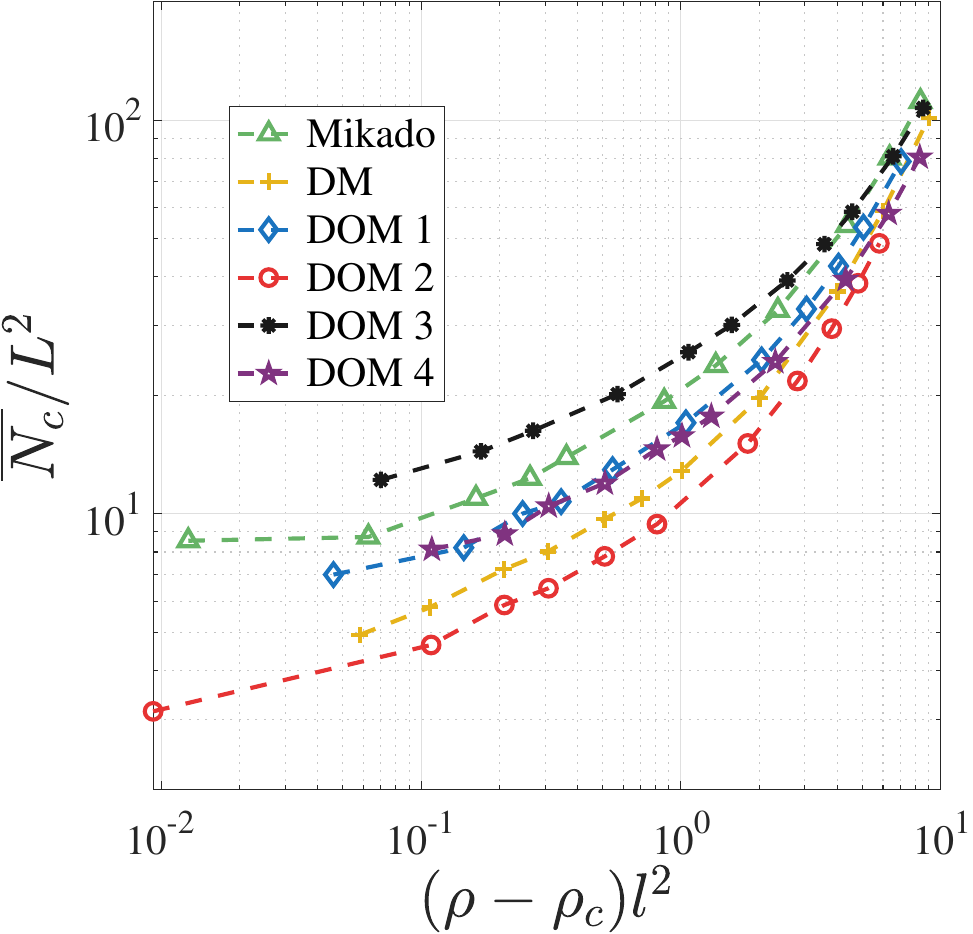}
         \caption{}
         \label{neper}
     \end{subfigure}
     \caption{The geometrical features of the networks near the percolation threshold: (a) network connectivity, (b) mean segment length, and (c) the mean number of segments per unit area.}
     \label{geof}
\end{figure}

Various geometrical features have been investigated near the percolation threshold and the results are shown in Figure \ref{geof} to give a better perspective on how different they are for the different network types in that region. The values of connectivity and average segment length remain constant just after percolation and then start to change, while $\overline{N_c}/L^2$ increases just after percolation as more clusters come into contact to the main percolated cluster with increasing density.

\section{Poisson's Ratio}

In order to calculate the value of Poisson's ratio, $\nu$, we used Hooke's law in 2D,
\begin{equation}
\begin{aligned}
    \varepsilon_{11} &= \frac{1}{E}(\sigma_{11} - \nu \sigma_{22}), \\
    \varepsilon_{22} &= \frac{1}{E}(\sigma_{22} - \nu \sigma_{11}), \\
    \varepsilon_{12} &= \frac{1 + \nu}{2E} \sigma_{12},
\end{aligned}
\end{equation}
where $\varepsilon_{ij}$ and $\sigma_{ij}$ are the $ij$-th components of the strain and stress tensors, respectively, and $E$ is the material's Young's modulus.

When we apply a purely longitudinal deformation in the $x_2$ direction, $\varepsilon_{11}$ and $\varepsilon_{12}$ will be zero; therefore, we have
\begin{equation}
    \nu = \frac{\sigma_{11}}{\sigma_{22}}.
\end{equation}
Since $\sigma_{11}$ and $\sigma_{22}$ are the normal stresses acting on surfaces whose normal vectors are perpendicular and parallel to direction of the uniaxial deformation, respectively, and the fact that we are applying small deformations to a square domain, we can write
\begin{equation}
    \nu=\frac{F_{\perp}}{F_{\parallel}}.
\end{equation}

\section{Self-Assembling Peptide Hydrogels}

\begin{figure}[H]
   \centering
     \begin{subfigure}[b]{0.99\textwidth}
         \centering
         \begin{overpic}[width=1\textwidth]{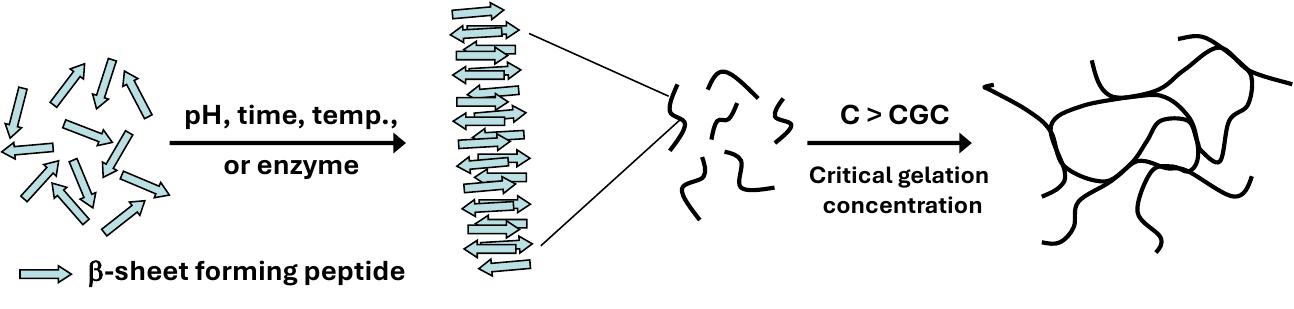}
         \end{overpic}
         \caption{}
         \label{PepFiber}
     \end{subfigure}
     \hfill \\
     \begin{subfigure}[b]{0.79\textwidth}
         \begin{overpic}[width=1\textwidth]{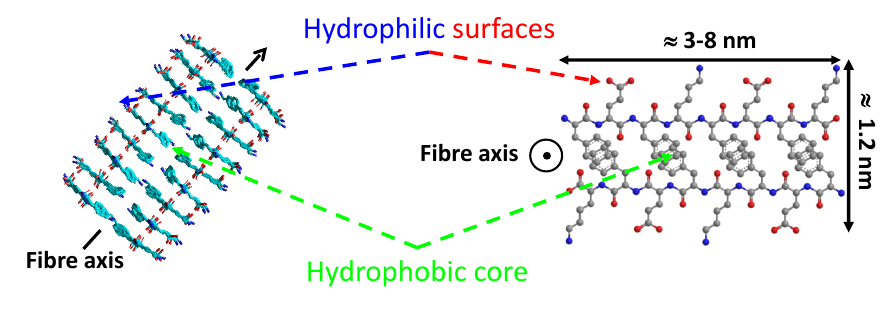}        
         \end{overpic}
         \caption{}
         \label{PepStructure}
     \end{subfigure}
     \hfill
     \begin{subfigure}[b]{0.2\textwidth}
         \begin{overpic}[width=1\textwidth]{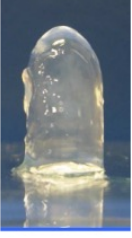}
         \end{overpic}
         \caption{}
         \label{Photo}
     \end{subfigure}
     \caption{Schematic representation of (a) the self-assembly and gelation pathway of β-sheet forming peptides and (b) the fibres formed (side and top views). (c) Photograph of  hydrogel.}
     \label{figH}
\end{figure}

The self-assembling peptide hydrogels (SAPHs) we used in our experiments are based on a design first developed by Zhang and coworkers [33-34], with alternating hydrophilic and hydrophilic residues. The design has been shown to result in peptides that self-assemble into antiparallel $\beta$-sheets with all the hydrophobic side groups being located on the same face of the $\beta$-sheet, while all the hydrophilic residues are located on the opposite face. It is hypothesised that, in water, two $\beta$-sheets come together to bury their hydrophobic faces resulting in the formation of relatively thick and rigid elemental fibres that, under appropriate conditions, associate/interact to form fibrillar networks, i.e. hydrogels. In Figure \ref{figH}, a schematic representation of the self-assembly and gelation process of these peptides is shown. The peptides form almost perfectly straight fibres. The fibres are sufficiently thick that thermal fluctuations can be neglected in our models.

\section{Materials and Methods}

\subsection{Materials:}
Peptides (over 95\% purity, HCl salts) were obtained from Biomatik LLC (Wilmington, Delaware, USA) and stored at -20°C.

\subsection{Hydrogel Preparation}
Peptides were dissolved in high-performance liquid chromatography-grade water and mixed using a vortexer. The pH was adjusted with NaOH, followed by vortexing and gentle centrifugation to remove bubbles. The overall charge of a peptide can be calculated to be
\begin{equation}
    \centering
\
|Z| = \sqrt{
\left( 
\sum_i \frac{N_i \cdot 10^{\text{p}K_{a_i}}}{10^{\text{pH}} + 10^{\text{p}K_{a_i}}} - 
\sum_j \frac{N_j \cdot 10^{\text{pH}}}{10^{\text{pH}} + 10^{\text{p}K_{a_j}}}
\right)^2}
\
 \label{Equation:4.1}
\end{equation}
where N\textsubscript{i/j} represent the numbers and pK\textsubscript{a\textsubscript{(i/j)}} are the pK\textsubscript{a} values of the basic (i: pK\textsubscript{a}  $>$ 7) and acidic (j: pK\textsubscript{a}$<$ 7) groups present, respectively [22].

The peptides in this study contain the following ionizable groups:
\begin{itemize}
    \item aarboxylic acid groups (COOH/COO\textsuperscript{-}) at the C-terminus (pKa 1.83 for F and 2.18 for K) and on glutamic acid side chains (pKa 4.25);
    \item amine groups (NH\textsubscript{3}\textsuperscript{+}/NH\textsubscript{2}) at the N-terminus (pKa 8.95) and on lysine side chains (pKa 10.53).
\end{itemize}

\begin{figure}[H]
\centering
\includegraphics[width=0.55\textwidth]{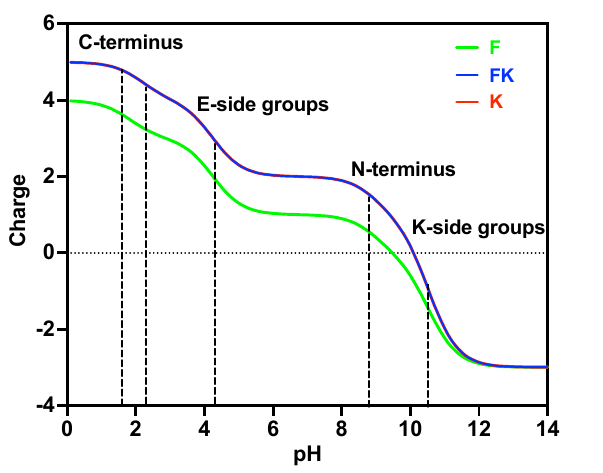}
\captionsetup{justification=centering}
\caption{The theoretical charge of the peptides as a function of pH.}
\label{pka}
\end{figure}

Figure \ref{pka} shows the theoretical charge of the peptides as a function of pH, with the pKa values of the different ionizable groups indicated. To ensure valid comparisons among the hydrogels, all were prepared to carry a charge of +2. Based on theoretical charge curves, peptides K and FK naturally align at +2 at pH 7, while peptide F, which has a charge of +1 at pH 7, was adjusted to pH 4.25 to achieve the target charge of +2.

\subsection{Transmission Electron Microscopy (TEM)}
TEM images were obtained from 10-fold diluted peptide hydrogels (10 mg/mL, charge +2). A 400-mesh carbon-coated copper grid (EM Resolutions, UK) was floated on a 10 μL droplet of the sample for 1 minute, followed by washing with water and staining with 1\% uranyl acetate for 30 seconds. After drying, the grid was imaged using a Talos L120C electron microscope.

\subsection{Oscillatory Rheology}
Rheological measurements were performed with a DHR-2 rheometer (20 mm parallel plate, 500 μm gap). An amplitude sweep (0.1–100\% strain, 1 Hz) determined the linear viscoelastic region (LVR), followed by a frequency sweep (0.2\% strain, 0.01–15 Hz) to analyse moduli. The storage modulus at 1 Hz and 0.2\% strain (in LVR) was recorded for each sample. Measurements were repeated three times for reproducibility.